# Enhancing Deterministic Freezing Level Predictions in the Northern Sierra Nevada Through Deep Neural Networks

Vesta Afzali Gorooh, Agniv Sengupta, Shawn Roj, Rachel Weihs, Brian Kawzenuk, Luca Delle Monache, F. Martin Ralph [a]

*Center for Western Weather and Water Extremes, Scripps Institution of Oceanography*

*University of California, San Diego*

*Corresponding author*: vafzaligorooh@ucsd.edu

ABSTRACT

Accurate prediction of the freezing level (FZL) is essential for hydrometeorological forecasting systems, with direct implications for runoff generation and reservoir management. In this study, we develop a deep learning–based postprocessing framework using the Unet convolutional neural network (CNN) architecture to refine the FZL forecasts from the West Weather Research and Forecasting (West-WRF) model. The proposed framework leverages reforecast data from West-WRF and FZL estimates from the California-Nevada River Forecast Center (CNRFC) to develop Unet models over the northern Sierra Nevada watersheds, such as the hydrologically critical Yuba-Feather watershed. We introduce two Unet model variants, Unet-log and Unet-GMM, that employ specialized loss functions beyond the standard benchmarks to enhance forecast skill. Unet-log utilizes the cosine of Error, and Unet-GMM uses Gaussian Mixture Model loss functions, to enhance FZL forecasts. Results show that Unet-based postprocessing reduces centered root-mean-squared errors by up to 20% and increases forecast–observation correlation by about 10% compared to raw West-WRF. Evaluation using the continuous ranked probability score (CRPS) for Unet-GMM further demonstrates consistent improvements across lead times. While performance fluctuates with forecast horizon, storm variability, and diurnal forcing, Unet-GMM and Unet-log consistently outperform the baseline. The models capture the spatiotemporal variability of the FZL across different elevations, mitigating biases from the West-WRF model. This novel deep learning–based postprocessing approach demonstrates a promising pathway for integrating machine learning into hydrometeorological forecasting and decision support within the Forecast Informed Reservoir Operations (FIRO) framework.

## 1. Introduction

Accurate estimates of the atmospheric freezing level (FZL)—defined in this study as the altitude above mean sea level (MSL) at which the atmospheric temperature reaches 0°C—play a key role in hydrometeorological forecasting, runoff prediction, and reservoir operations. This is particularly important in the context of Forecast Informed Reservoir Operations (FIRO), a modern water management strategy that incorporates atmospheric and hydrologic forecasts into real-time reservoir decisions (Jasperse et al. 2020; Ralph et al. 2019). In the Northern Sierra Nevada Mountains, where complex topography and rapidly changing atmospheric conditions interact, landfalling atmospheric river events (AR; e.g., Zhu and Newell 1994, 1998; Ralph et al. 2004, 2005) play an integral role in shaping the water year (Neiman et al. 2008; Dettinger 2011; Dettinger and Cayan 2014). Dynamical models such as the Weather Research and Forecasting (WRF) model often produce FZL forecasts with significant biases and uncertainties. Observations and models have shown the complexity and variability of the FZL in this mid-latitude mountainous region (Sumargo et al. 2020; Henn et al. 2020; Minder and Kingsmill 2013; Cannon et al. 2017). For instance, within the Yuba-Feather watershed, the FIRO assessments (Preliminary and Final Viability Assessments, Yuba-Feather PVA & FVA, CW3E, UC San Diego) highlight systematic forecast limitations in modeling snow-rain transition processes, particularly under the western configuration of WRF, West-WRF. These assessments identify snow-level forecast uncertainty, arising from misrepresentation of FZL and melting-layer structure, as a key operational risk affecting reservoir pre-release strategies at Lake Oroville and New Bullards Bar (Yuba-Feather PVA & FVA, CW3E UC San Diego). While FZL alone is an insufficient proxy for surface precipitation phase, it remains the essential gridded thermodynamic forecast variable, available to water managers. Accurate FZL forecasts, therefore, directly reduce uncertainty in the rain–snow partitioning and the downstream streamflow predictions that underpin FIRO decision-making.

Building on this operational context, Sumargo et al. (2020) quantified the consequences of FZL forecast error for the same watershed, specifically Lake Oroville and New Bullards Bar Reservoirs, that FZL forecast errors can lead to significant misrepresentations of storage, consuming 2–10% of flood pool volume for every inch of precipitation, depending on local conditions. They further demonstrated that sensitivity peaks when the FZL occupies the mid-elevation band characteristic of these watersheds, reinforcing the need to reduce error precisely where operations are most vulnerable. Their findings underline the sensitivity of the Feather and North Fork Yuba River watersheds to FZL variability and emphasize the motivation for our study's focus on improving FZL forecasts used within FIRO workflows. Other recent studies have also highlighted the nonlinear relationship between the FZL and consequential hydrological responses, particularly in terms of runoff variability. Qin et al. (2019) analyzed the FZL and summer runoff in the Hotan River Basin, China, using nonlinear decomposition techniques and hybrid prediction models. The study revealed significant inter-annual and inter-decadal FZL cycles, emphasizing the importance of accounting for nonlinear interactions in hydrological modeling. These results underscore the critical role of advanced data-driven approaches in improving FZL-based hydrological forecasting, which has implications for flood risk assessment and water resource management.

Hatchett et al. (2017) found that FZL can vary from 1,000 to 3,500 m, sometimes within a single storm. This variability presents notable difficulties for forecasters and reservoir managers. Henn et al. (2020) reported FZL displacement by comparing ZQ50, the model-estimated altitude where the fraction of frozen hydrometeors drops to 50%, from three numerical weather prediction models, against radar-derived bright band observations (White et al. 2002; White et al. 2010) from vertically pointing snow level profilers across California. Focusing on a 4-month period ending in March 2017, their analysis using the West-WRF model (Martin et al. 2018) revealed a mean displacement of 207 meters, with an interquartile range of 130 meters (140–270 m). These findings emphasize that freezing level and snow level displacement vary significantly with model type, location, and evaluation methodology, highlighting the need for region-specific model improvement. Dynamical model forecast errors arise from multiple sources, including initial conditions, boundary conditions, and structural model deficiencies (Delle Monache et al. 2013). The chaotic nature of atmospheric dynamics further amplifies these errors over time, limiting forecast skill and operational reliability. These limitations impair the accuracy of operational forecasts and exacerbate risks associated with extreme precipitation and unexpected icing or runoff events.

Traditional postprocessing techniques, such as bias correction methods, Kalman filters (e.g., Delle Monache et al. 2011), linear regression-based postprocessing techniques, and Model Output Statistics (MOS; Glahn and Lowry 1962), have been widely employed to systematically improve the accuracy in numerical weather prediction (NWP) forecasts by establishing statistical relationships between raw model outputs and observations (e.g., Wilks and Hamill 2007; Taillardat et al. 2016). However, these methods are inherently struggling to capture the complex, highly non-linear interactions governing the atmosphere, limiting their ability to accurately model rapid changes in variables like FZL, especially in regions with rapid changes and strong orographic effects (Vannitsem et al. 2021). Deep learning postprocessing approaches offer a promising alternative to traditional statistical methods for improving NWP forecast skill. In particular, recent advances using deep neural networks have demonstrated substantial improvements in the postprocessing of hydrometeorological variables, such as temperature (e.g., Alerskans et al. 2022; Cho et al. 2022), precipitation (e.g., Ghazvinian et al. 2021, 2022, 2024; Hu et al. 2023; Badrinath et al. 2023, Li et al.

2022), sea ice (Palerme et al. 2024), wind speed (Sun et al. 2024) and integrated vapor transport (IVT; Chapman et al. 2019, 2022). These deep learning methods, particularly convolutional neural networks (CNNs) with Unet architectures, leverage customized loss functions to capture fine-scale spatial features outperforming traditional approaches in reducing forecast errors (Pan et al. 2019; Zhang et al. 2023). While the success of deep learning for precipitation and related variables is promising, few studies have directly addressed the more complex challenge of FZL postprocessing in mountainous terrain, where observational and modeling uncertainties remain high (Sumargo et al. 2020; Henn et al. 2020).

Inspired by these insights, this study addresses a critical gap in operational hydrometeorology: the need for more accurate, high-resolution postprocessed FZL forecasts in mountainous watersheds such as the Yuba-Feather watershed, a region with a long history of catastrophic floods and a primary basin under the FIRO (Talbot et al. 2019; Jasperse et al. 2015, 2020). Despite the clear operational significance of FZL forecast errors for flood risk and reservoir management in California (Sumargo et al. 2020), few studies have directly targeted FZL postprocessing with advanced data-driven approaches. Our central research question is: can a deep learning-based postprocessing framework reduce errors in regional model-based FZL forecasts? To address this, we: (1) explicitly target the freezing level, an operationally critical variable, for postprocessing reginal dynamical model using a Unet approach tailored to the complex topography of the Sierra Nevada's watershed's; (2) implement advanced loss functions to capture FZL forecast errors; and (3) demonstrate substantial performance improvements at operationally relevant lead times. These innovations directly address identified limitations and clearly differentiate our study, emphasizing its practical and scientific contributions.

The remainder of this paper is organized as follows: section 2 describes the data sources utilized in our study, while section 3 details the proposed postprocessing methodology over the study area. Section 4 presents the model validation and performance assessment against observational datasets.

## 2. Data Sources

### *a. Dynamical Model Forecasts*

In this study, FZL forecasts from the West-WRF reforecast (Cobb et al. 2023) were used for the postprocessing model development. The reforecast is a 33-year-long dataset over the Western U.S. with diverse atmospheric and hydrological conditions. It provides insights into representative high-resolution atmospheric dynamics during the months of December through March. This dataset is a valuable source of information for data-driven algorithm development, such as deep learning postprocessing methodologies and in-depth process-based studies (Cobb et al. 2023). The West-WRF reforecast is based on WRF version 4.0.1 and is initialized daily at 0000 UTC using the Global Ensemble Forecast System Version 10, which refines the global model's coarse 0.5-degree resolution down to high-resolution 9 km and 3 km grids. The reforecast employs 60 vertical levels with an adaptive time-stepping approach, optimizing numerical stability across varying atmospheric conditions. The 3-km domain used in this study provides high-resolution forecasts for detailed analysis for 120 hours, while the 9-km domain offers extended lead times of up to 168 hours. The forecast files were generated every 3 hours, allowing for comprehensive temporal resolution in the training dataset. Nine winter seasons (December 2010 – March 2019) were used for the postprocessing model training and evaluation, capturing the period when FZL fluctuations are most critical. Since FZL is not directly provided in the reforecast product, it is derived by scanning downward through each vertical column (in each grid cell) to locate the first level where the temperature crosses 0 °C. If no crossing is found—indicating that the entire column is below freezing—the FZL is set to the model terrain height, representing a lower bound below which the FZL would occur.

### b. *Ground Truth / Reference*

7

## 3. Methodology and Experimental Design

### a. *Preprocessing West-WRF reforecast and CNRFC data:*

The West-WRF reforecast dataset and the CNRFC FZL analysis files were harmonized through a series of preprocessing steps to facilitate a robust postprocessing analysis. Given that the West-WRF reforecasts are at a native 3-km resolution, while the CNRFC FZL analyses are available on a 4-km HRAP grid, a careful regridding procedure was essential to ensure spatial consistency.

1. *Th*e CNRFC files, which serve as the ground truth for FZL analyses, were retained without modification. The HRRR Digital Elevation Model (DEM) was rescaled to the 4-km CNRFC HRAP grid.

2. Any CNRFC FZL value that fell below the corresponding elevation from the 4-km DEM was set equal to the DEM elevation, ensuring that freezing level values did not drop below the physical terrain.

3. Within each West-WRF reforecast file, the first occurrence of a 0°C crossing was identified for each grid cell, starting from the top of the model domain. In cases where no 0°C crossing was observed, indicating that the entire atmospheric column remained below freezing, the FZL was assigned the value of the terrain height.

4. The computed 0°C crossing elevations from the West-WRF reforecast (originally on the 3-km grid) were gridded (bilinear interpolation) to align with the 4-km HRAP grid of the CNRFC FZL analysis files.

5. The matched datasets used for training the postprocessing models were generated at six-hourly intervals (0000, 0600, 1200, and 1800 UTC), aligning with the temporal resolution of the reference CNRFC dataset. The input variables included the regridded FZL derived from West-WRF reforecasts (step 4) and static geographical information, including, DEM, latitude, and longitude. Models were trained exclusively on data from the winter seasons (December–March) of 2010–2015 and 2017–2019.

These preprocessing steps ensured consistency between the West-WRF and CNRFC datasets, supporting the development of the deep learning postprocessing framework described in Section 3.2.

### *b. Unet Framework*

The CNN with Unet architecture, introduced by Ronneberger et al. (2015), has been widely used and consists of a contracting path to capture context and a symmetric expanding path that enables precise localization. Figure 1 shows a general Unet adapted to process West-WRF model outputs and predict the FZL in an effort to improve accuracy. The encoder systematically reduces the spatial dimensions of the input data (West-WRF FZL) while increasing the number of feature channels. This is achieved through convolutional layers with kernels applied sequentially. Each convolutional layer is followed by a Leaky rectified linear unit (LeakyReLU) activation function to introduce non-linearity. After every convolution set, a max-pooling operation is utilized to downsample the feature maps effectively and halve the spatial dimensions, allowing the model to capture hierarchical features at multiple scales.

The decoder reconstructs the spatial dimensions by upsampling the feature maps. Upsampling is performed using transposed convolutions followed by standard convolutions. At each decoder stage, the upsampled feature maps are concatenated with the corresponding feature maps from the encoder via skip connections. A. The final layer of the Unet employs a 1×1 convolution to map the multi-channel feature maps to the desired output (here, postprocessed FZL). This study implemented the Unet architecture (detailed in Section 3d) to refine the FZL from West-WRF model outputs. The input data is formatted similarly to image (latitude, longitude) data, allowing the network to leverage spatial patterns and correlations inherent in atmospheric phenomena.

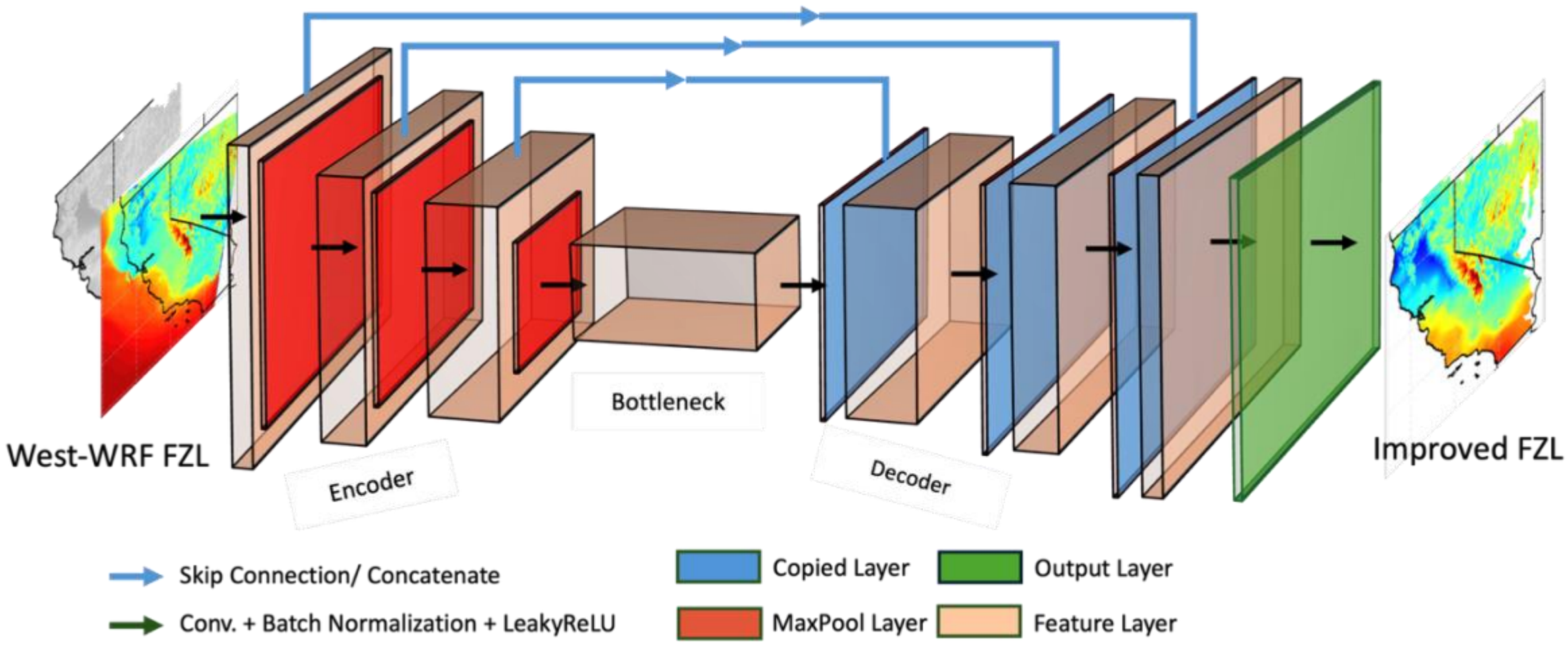

Figure 1. Schematic of the Unet architecture for freezing level postprocessing.

### *c. Loss Functions*

First, we train a Unet model with root mean squared error (RMSE) loss function (Unet-RMSE), and a linear regression model (hereafter Regression) used as benchmarks for the two proposed Unet models that use customized loss functions. The Regression model employs a linear approach using Stochastic Gradient Descent (SGD) to learn the relationship between input features and FZL forecasts at different lead times. It is trained incrementally for each forecast lead time using mini-batches, enabling efficient scaling to large datasets while maintaining low memory usage. Each three-dimensional (3D) input sample is normalized to ensure consistent scaling across different variables and is treated each grid cell as an independent training sample. The model learns to map the 3D input to corresponding reference FZL values, with separate models trained for different forecast lead times to capture the varying predictive relationships at different temporal horizons. This provides a straightforward baseline to assess the added value of the convolutional deep learning architectures. The two customized loss functions are: (1) the Log-Cosh error function, which is a smooth approximation of mean absolute error (MAE) and combines the advantages of mean square error (MSE) and MAE, and (2) a Gaussian Mixture Model (GMM) loss function that considers the distribution of the prediction. The details of each of these loss functions are provided below.

### Log-Cosh Error Loss Function

The Log-Cosh loss function is defined as the logarithm of the hyperbolic cosine of the prediction error (Jadon et al. 2024). It provides a robust alternative to traditional loss functions by smoothly approximating MAE and combining the benefits of MSE and MAE. This effectively handles both small and large errors, which is particularly useful in meteorological data, where outliers can significantly affect model performance.

The loss function is defined as

$$Loss = \frac{1}{N}\sum_{i=1}^{N} log(cosh(\widehat{y_i} - y_i) \qquad (1)$$

Where $N$ is the number of samples, $y_i$ is the true value for the $i^{th}$ sample and $\widehat{y_i}$ is the predicted value.

### Gaussian Mixture Model Loss Function

Empirical analysis of the observed FZL height distribution over northern Sierra Nevada reveals a clear bimodal structure (Figure 1). The histogram indicates the presence of two distinct modes, which are well-captured by a two-component GMM. This empirical evidence supports use of a bimodal GMM for the loss function, as it provides a substantially improved fit compared to a unimodal Gaussian or gamma distribution and more accurately reflects the observed FZL variability. The GMM loss function models the prediction error as a mixture of two Gaussian distributions, allowing the network to capture complex error distributions, by maximizing the likelihood of observed values under this mixture (Bishop 1994). We parameterize the probability density function *F(x)* as:

$$F(x) = \pi_1 . N(x, \mu_1, \sigma_1^2) + \pi_2 . N(x, \mu_2, \sigma_2^2) \qquad (2)$$

and minimize the negative log-likelihood of the observations under *F(x)* (see Eq. 4). Where $\pi_1$ and $\pi_2$ are the mixing coefficients such that $\pi_1 + \pi_2 = 1$ and $N(x, \mu_i, \sigma_i^2)$ is the Gaussian distribution of x:

$$N(x, \mu_i, \sigma_i^2) = \frac{1}{\sqrt{2\pi\sigma_i^2}} exp(-\frac{(x-\mu_i)^2}{2\sigma_i^2}) \tag{3}$$

The loss function is then defined as the negative log-likelihood:

$$Loss = -\frac{1}{N}\sum_{i=1}^{N} log(F(x_i)) \tag{4}$$

*d. Model training and configuration*

The Unet models are trained separately using the Log-Cosh and GMM loss functions (hereafter Unet-log and Unet-GMM, respectively). We trained different models for each lead time to better capture temporal dynamics. The Unet architecture employed in this study is a compact encoder–decoder network (as shown in Figure 2), consisting of one downsampling level, a bottleneck with dilated convolutions, and a symmetric upsampling path with skip connections. The datasets include West-WRF model outputs (as input to the Unet) and observed CNRFC FZL measurements (as ground truth). The Unet architecture is summarized in Table 1 and was kept identical across experiments; the only difference among models was the choice of loss function (e.g., log-cosh or GMM). The layers are connected in sequence from the top of the table to the bottom. The table shows each component of the encoder–decoder structure, including the number of layers, kernel sizes, activation functions, and output channels. We employed a 3×3 convolutional kernel throughout the network, a standard choice in Unet implementations that balances the ability to capture fine-scale spatial features with computational efficiency. Training involves optimizing the network parameters to minimize loss functions using stochastic gradient descent with Adam optimizer with a learning rate of $1\times10^{-3}$, LeakyReLU activations ($\alpha = 0.2$), Glorot uniform kernel initialization (Glorot and Bengio 2010), batch normalization after each convolution, and early stopping with a patience of 20 epochs to prevent overfitting. The batch size was set to 100, and the networks were trained for up to 100 epochs. These design choices ensured the model's ability to capture FZL dynamics while maintaining stable training performance.

The training period consisted exclusively of winter seasons (December–March) from 2010–2015 and 2017–2019, reflecting the importance of freezing level predictions during the cool season for determining precipitation phase. One winter season was withheld for validation, and the remaining winters from the training period were used for training. This one-year cross-validation approach ensured that the validation set represented an independent winter and allowed for robust hyperparameter optimization. For independent testing, we used two additional winter seasons not included in training the model: December 2015–March 2016 and December 2016–March 2017. This setup ensured that the model wsas evaluated under distinctly different climatic conditions while maintaining consistency with the winter-only training focus. The 2016–17 season brought record-breaking precipitation across California, while the 2015–16 season remained unexpectedly dry despite a strong El Niño event, highlighting the role of internal atmospheric variability in modulating precipitation (Kumar and Chen 2017; Luna-Niño et al. 2025; Siler et al. 2017). Appendix A provides maps of accumulated precipitation for both winters, as well as a difference map to emphasize the spatial variability between the two seasons.

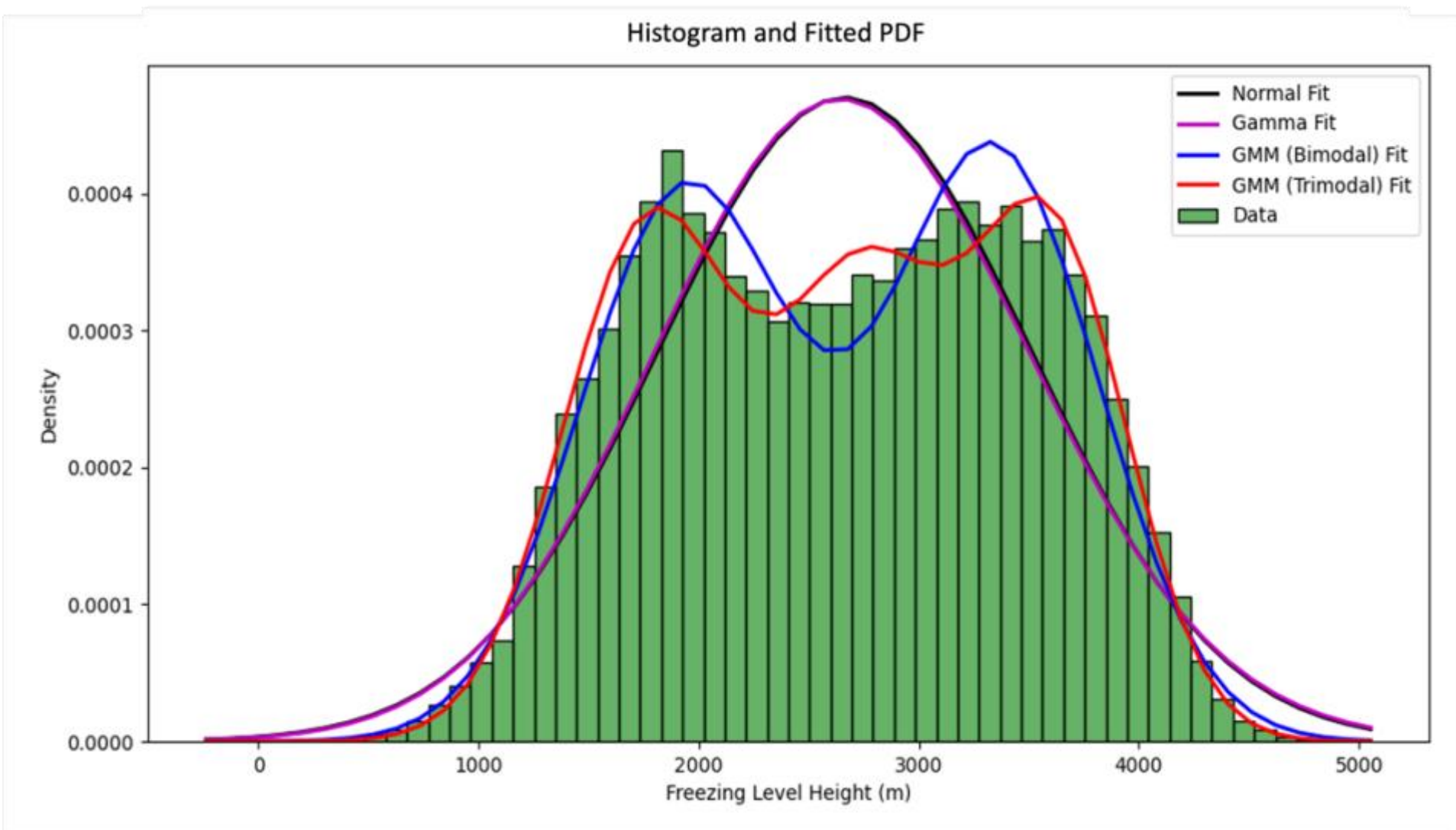

Figure 2. Visualization of empirical probability distribution of observed FZL heights within the northern Sierra Nevada during the study period (2010-2019) and fitted curves: normal, gamma, bimodal Gaussian Mixture Model (GMM) and trimodal GMM. The histogram (green bars) reveals a bimodal distribution, well-captured by a two-component GMM (blue curve).

Table 1. Summary of the Unet architecture used in this study. (*) shows the output layer for deterministic postprocessing models (Unet-RMSE, and Unet-log) with 157,729 trainable parameters, (**) show the GMM components for Unet-GMM model with 157,861 trainable parameters.

| Layers | Filter Size / Stride | Channels | Activation | Notes |
|---|---|---|---|---|
| Input | - | 4 | - | 2D input field on 4-km grid Static DEM, latitude, longitude + West-WRF FZL predictors |
| Encoder | 3 × Conv2D (3×3, stride 1, same padding) | 32 | LeakyReLU | MaxPooling2D (2×2) and Batch normalization applied |
| Bottleneck | 3 × Conv2D (3×3, stride 1, same padding) with dilations (1, 2, 4) | 64 | LeakyReLU | Skip connections summed |
| Decoder | Conv2DTranspose (2×2, stride 2) + 3 × Conv2D (3×3, stride 1) | 32 | LeakyReLU | Concatenated with encoder skip, Conv2DTranspose (2×2) applied |
| Output* | Conv2D (1×1) | 1 | Linear | Postprocessed FZL output for Unet-RMSE, and Unet-log |
| Output Branch 1 | Conv2D (1×1) | 4 | Linear | Mixture density outputs ($\mu_1$, $\sigma_1$, $\mu_2$, $\sigma_2$) |
| Output Branch 2 | Conv2D (1×1) | 1 | Sigmoid | Probability weight $\pi_1$ ($\pi_2$=1- $\pi_1$) |
| Output** | Concatenate Output Branch 2 with Output Branch 1 | 5 | - | Combined GMM parameters Unet-GMM |

*e. Case Study Area*

Figure 3 shows the study domain encompassing the Yuba-Feather River watershed, with the topographic map including the North Fork Feather, East Branch North Fork Feather, Middle Fork Feather, and Upper Yuba basins. The Yuba River, originating in the Sierra Nevada, is one of California's historic rivers, with its main tributaries being the North, Middle, and South Yuba Rivers. The Feather River, the largest in the Sierra Nevada, begins in the northern Sierra Valley and flows into the Sacramento River. For this proof-of-concept study, we selected the Yuba-Feather watershed, one of the most hydrologically complex and operationally important watersheds in the northern Sierra Nevada, due to its challenging topography, orographically enhanced precipitation, and frequent interactions with ARs. With hydropower and flood-control reservoirs, the basin requires accurate FZL predictions to improve flood mitigation strategies and optimize water storage efficiency. Given its highly variable precipitation regimes and sensitivity to FZL fluctuations, this region provides an ideal case study for evaluating machine learning-based postprocessing techniques aimed at reducing FZL forecast biases.

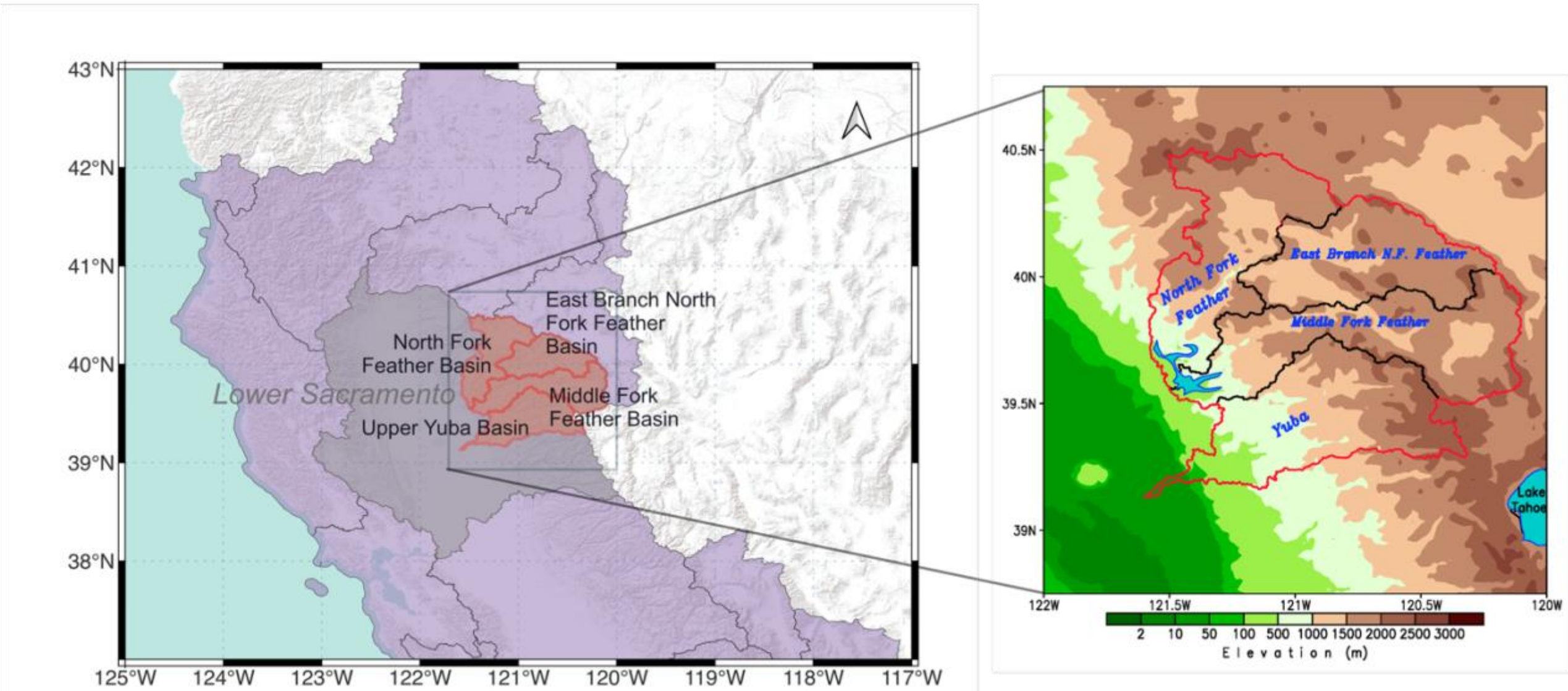

Figure 3. The case study domain and the elevation map (in the inset) of the Yuba Feather River watershed including North Fork Feather, East Branch North Fork Feather, Middle Fork Feather, and Upper Yuba basins.

## 4. Results and Discussions

In this section, we present the results of the Unet-based postprocessing models over two winter seasons independent of the training period: a wet year (2016-17) and a relatively dry year (2015-16). Since the reference data, CNRFC, and West-WRF (hereafter WRF) are deterministic, we evaluate the Unet-GMM model using the expected value of the predicted distribution, as defined in Equation (2). We begin by presenting the time series of FZL forecast performance, followed by the spatial distribution of skill metrics comparing the original WRF forecasts and the postprocessed forecasts relative to CNRFC observations across different lead times. Finally, we assess the models' skill across varying terrain elevations to understand their performance in complex topography.

### *a. Performance of freezing level forecasts postprocessed with deep learning*

Figures 4 and 5 illustrate the performance of the Unet postprocessing models (Unet-RMSE, Unet-log, Unet-GMM) and the Regression benchmark as a function of lead time against CNRFC observational target over the Yuba-Feather watershed for the winters of 2016–17 and 2015–16, respectively. The results are computed using winter season samples aggregated over grid points covering the study area, offering insight into the temporal evolution of forecast skill. All three Unet models exhibit noticeable improvements in FZL forecasts in both the Pearson correlation coefficient (hereafter correlation) and centered root mean square error (CRMSE) when compared to the raw WRF forecasts during the relatively wet 2016–17 season (Figure 4). Definitions of these verification metrics are provided in Appendix B. The Unet-based methods (blue, red, and grey lines) consistently achieve higher correlation and lower RMSE values than the baseline forecasts (black line), indicating improved forecast accuracy. During the 2016–17 winter, the average CRMSE across all lead times decreased from approximately 360 m in the raw WRF forecasts to 340 m with Unet-RMSE and further to 300 m with both Unet-GMM and Unet-log. Similarly, the correlation improved from an average of 0.85 in WRF to 0.88 with Unet-RMSE and further to 0.90 with the Unet-GMM and Unet-log models. These results indicate that while Unet-RMSE effectively reduces errors compared to the raw forecasts, the probabilistic models (Unet-GMM and Unet-log) provide additional gains in both accuracy and consistency. Depending on the lead time, these Unet approaches enhance forecast skill by approximately 5% to 20%, with performance naturally degrading from Day 1 to Day 3 and beyond. Temporal variability of the correlation shows that the Unet-log and Unet-GMM methods perform best, reaching correlation values above 0.95 at short lead times (6–24 hours), whereas the raw WRF rarely exceeds 0.93. This temporal variation may be partially influenced by more FZL values being at the surface during the 12Z forecasts, which results in closer agreement between model output and observations. In addition, the noticeable fluctuations in performance metrics across lead times and seasons can be attributed to the interaction of synoptic-scale variability, storm intensity, and diurnal forcing, which affect how rapidly errors accumulate in WRF forecasts and how effectively the Unet models can correct them. These factors lead to periods where postprocessing yields larger relative gains, while at other lead times the improvements are less pronounced. Although the relative improvement is less pronounced during winter 2015-16, the overall improvement in FZL forecasts from the Unet models across all lead times remains evident (Figure 5). The correlation remains high, ranging from 0.87 to 0.97, while the CRMSE reductions vary from 5% to 20%—demonstrating that, under both wet and dry winter conditions, Unet-based postprocessing models consistently improve forecast accuracy. At most of the lead times, Unet-Log and Unet-GMM (blue and red) outperform Unet-RMSE, reinforcing their superior skill in postprocessing WRF FZL forecasts.

In both winters, the raw WRF and Regression forecasts exhibit higher CRMSE, ranging from 275 m to 460 m, particularly at longer lead times, when forecast uncertainty tends to increase. The FZL improvements by Unet models are consistent but more pronounced during the winter of 2016–17, where complex precipitation dynamics and enhanced moisture transport challenge dynamical models such as WRF. Conversely, in the 2015–16 winter season, the relative improvement from postprocessing is less prominent, suggesting that synoptic-scale variability and precipitation frequency influence the effectiveness of machine learning-based correction methods. The improvements in correlation and CRMSE for Unet-GMM and Unet-log, compared to raw WRF forecasts, remain robust at most lead times for both seasons, particularly within the first 72 hours. Beyond 72 hours, while improvements persist, the interquartile ranges of errors begin to overlap more frequently, reflecting greater uncertainty at longer lead times. The temporal variability of the correlation (Figures 4a and 5a) and CRMSE (Figures 4b and 5b) suggests that Unet-based methods perform best at most of the lead times. This improvement is consistent across both wet and dry years, underscoring the robustness of the Unet-based postprocessing in various atmospheric regimes. Notably, while all three Unet frameworks outperform the raw WRF forecasts, Unet-log and Unet-GMM consistently yield the best results, highlighting the impact of different loss functions and distributional assumptions on postprocessing efficacy. Given this consistent trend, we will focus on the verification skill results from the Unet-GMM and Unet-log in the remainder of the paper.

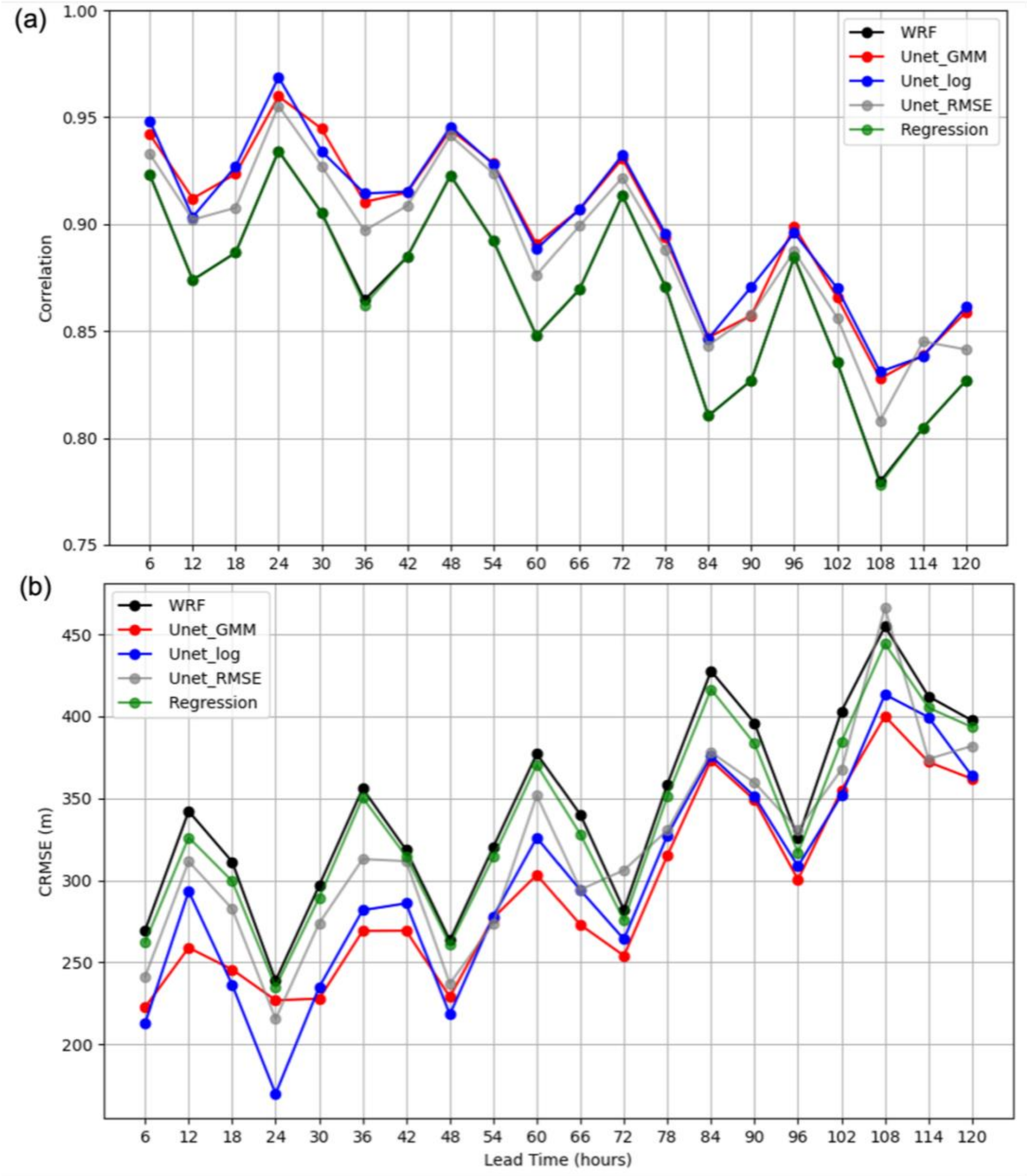

Figure 4. Performance metrics of 6-hourly WRF FZL and Unet postprocessing (Unet-RMSE, Unet-GMM, and Unet-log) forecasts during December 2016 - March 2017 over the Yuba-Feather Watershed: (a) Pearson correlation coefficient and (b) RMSE versus CNRFC observations as a function of lead time (hours).

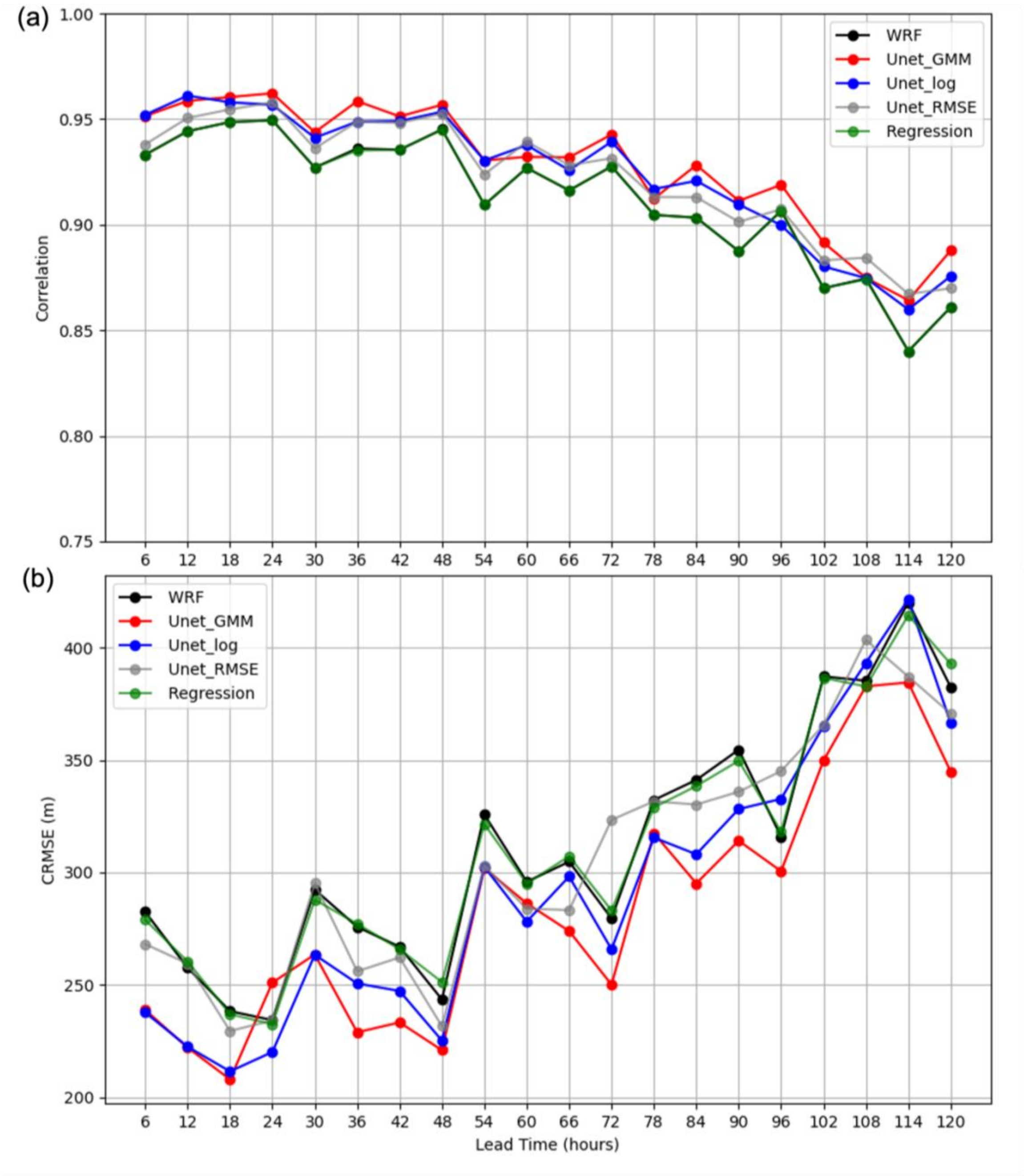

Figure 5. Same as Figure 4, but for December 2015 - March 2016.

Figure 6 illustrates the spatial distribution of RMSE and correlation coefficients for Unet-GMM and Unet-log FZL forecasts, as well as the baseline WRF forecasts with respect to CNRFC observations for the winter of 2016-17 at a 2 Day lead time. WRF exhibits high errors, particularly in the central and eastern regions of the Yuba-Feather watershed, with RMSE values ranging between 230 and 310 m (Figure 6a). These areas correspond to higher terrain, where FZLs are more frequently at the surface, potentially contributing to larger errors due to mismatches in vertical structure. However, both deep learning models, Unet-GMM and Unet-log, significantly reduce RMSE over these regions to below 200 m in most of these areas (Figure 6b-c). Both Unet models also reduce the errors in the eastern parts of the watershed, a region characterized by complex terrain and high elevations (above 1,500 m), where FZL forecasting remains challenging. Between the two postprocessing models, Unet-log reduces RMSE by an additional 10–15% compared to Unet-GMM for this lead time over most of the basin. The corresponding correlation maps (Figure 6d–f) display strong agreement between the FZL forecasts and the observed CNRFC FZL, particularly in the western part of the watershed, where correlation values consistently exceed 0.85–0.90 across most grid points. Compared to WRF, which shows weaker correlations in certain areas in the eastern basin, both Unet models maintain a slightly higher spatial coherence with observations. Notably, the Unet postprocessing models achieve the lowest (highest) overall error (correlation), confirming that they not only enhance accuracy but also preserve spatial structure in FZL forecasts.

The results for the winter of 2015-16 exhibit elevated RMSE in the southwestern part of the basin, as shown in Figure 7. WRF forecasts show RMSE values exceeding 280 m in most areas, while Unet models reduce errors, particularly in the high-elevation eastern regions. Unet-GMM and Unet-log demonstrate notable improvements in FZL forecasts across the entire basin, especially in high altitudes (Figure 7b). This pattern aligns with the characteristics of a dry season, which has fewer precipitation events and generally higher freezing level elevations in the lower-elevation areas of the basin compared to a wetter season (2016-17 winter season). Correlation maps (Figure 7d–f) for the 2015-16 winter confirm that both deep learning models enhance spatial agreement with observations (reducing the RMSE) while maintaining correlation values consistently above 0.95 across most areas. These improvements are particularly evident in regions where WRF exhibits large biases, further underscoring the potential of machine learning approaches for improving numerical weather prediction model outputs. We have also examined other lead times and found that the improvement provided by the Unet postprocessing models remains consistent across most forecast horizons, further supporting their robustness in reducing errors. In the next section, we present the evaluation of a case study during an extreme AR event to demonstrate the effectiveness of the postprocessing models under extreme weather conditions.

As shown in Figures 6 and 7, the performance of the Unet models varies between seasons in different regions of the watershed. In the wetter 2016-17 winter, the greatest RMSE reduction occurs in the lower-elevation western areas (~100–1,500 m), where Unet-log outperforms other models. However, in the drier winter of 2015-16, the Unet models primarily reduce errors in the higher-elevation eastern regions (above 1,500 m), suggesting a potential dependency of model performance on seasonal precipitation regimes and terrain complexity. In Section 4.3, we further evaluate the performance of the FZL forecasts by considering the model errors based on elevation. This will allow us to assess whether the improvements are consistent across different terrain levels or whether the model's effectiveness varies based on the underlying topography and meteorological conditions. By doing so, we aim to better understand the limitations and strengths of deep learning-based postprocessing in operational forecasting.

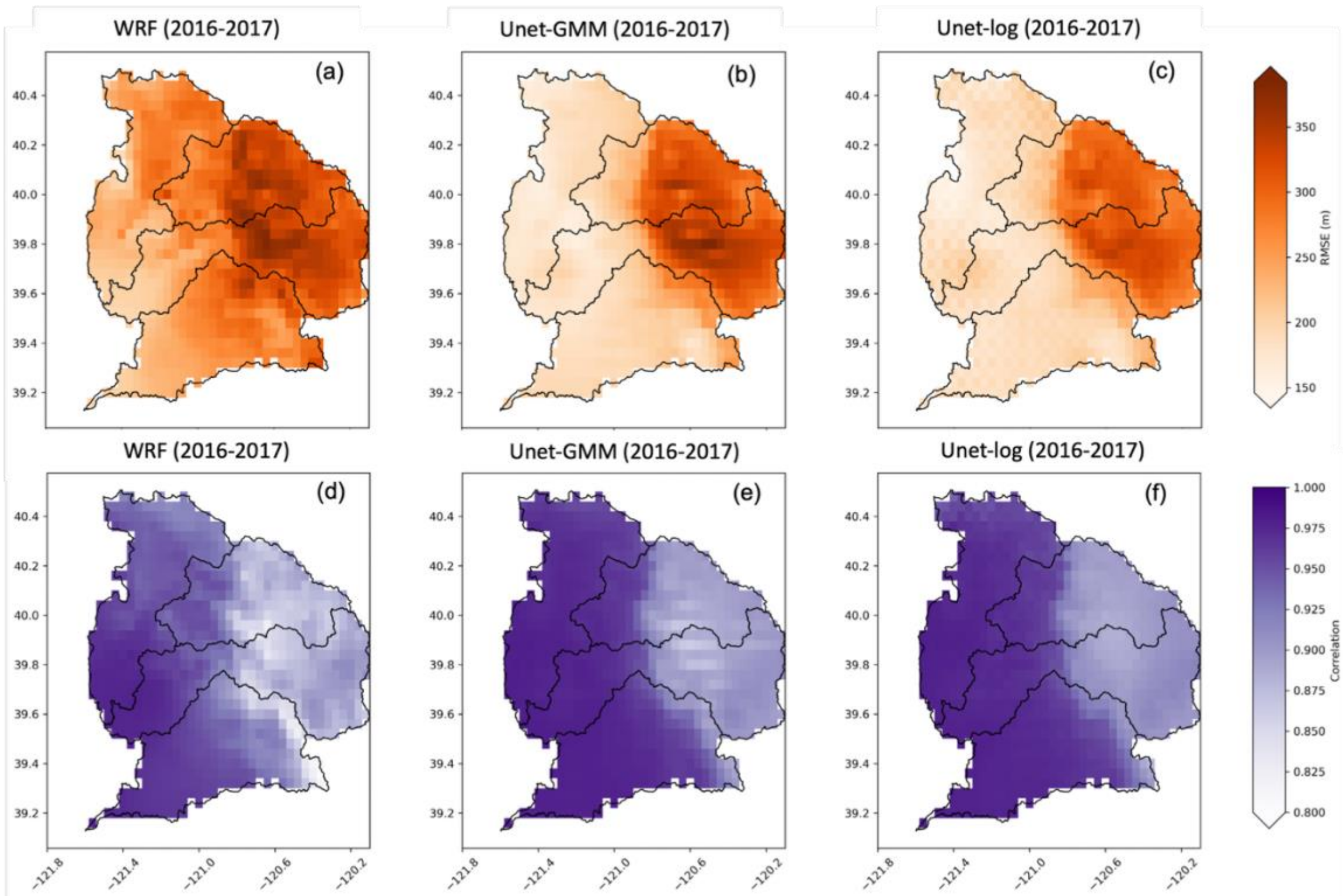

Figure 6. RMSE (a-c) and correlation (d-f) of 48-hour lead time FZL forecasts during winter 2016-2017 for (a and d) WRF, (b and e) Unet-GMM, and (c and f) Unet-log against observations from CNRFC.

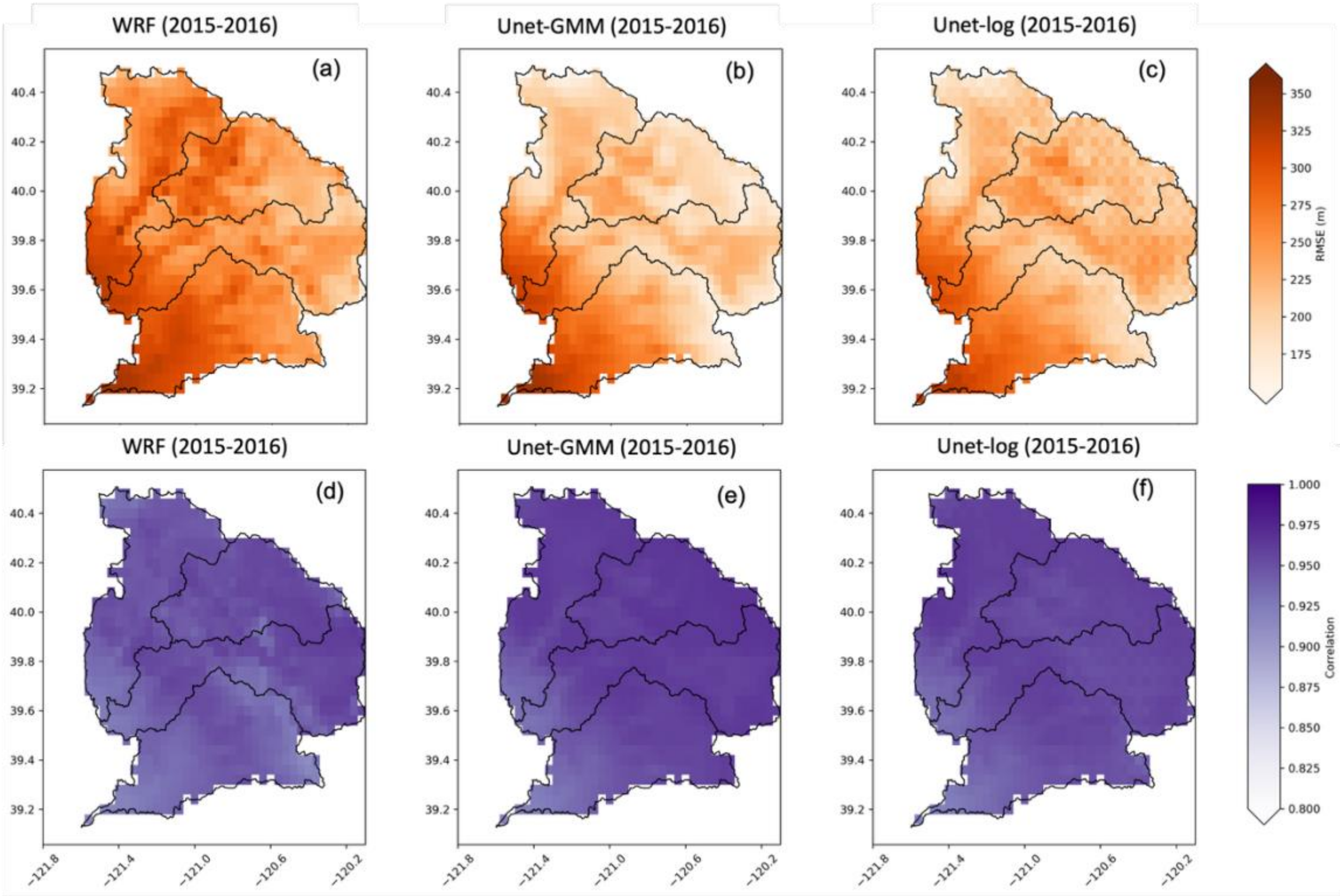

Figure 7. Same as Figure 6 but for winter 2015-2016.

To complement the CRMSE and correlation analyses presented earlier, we evaluate the continuous ranked probability score (CRPS) for the predictive distributions from Unet-GMM and the forecasts from Unet-log and WRF, even though the goal of this study is deterministic postprocessing (for deterministic forecasts, CRPS equals MAE). As shown in Figure 8, in both winters, CRPS increases with lead time for all methods. Still, Unet-GMM consistently attains the lowest values, Unet-log is intermediate, and WRF is the highest. During the wet winter of 2016–17 (Figure 8a), Unet-GMM reduces CRPS by roughly 10–35% relative to WRF, with the largest gains at medium–long leads (~48–120 h); Unet-log also improves on WRF but to a lesser extent. The dry winter of 2015–16 (Figure 8b) shows overall smaller CRPS. Yet, the same ranking persists, including long-lead spikes where Unet-GMM retains the best performance. These results confirm that the probabilistic Unet-GMM yields more accurate FZL forecasts and that CRPS is a valuable complement to the CRMSE plots by summarizing distributional accuracy (and MAE for deterministic models) across lead times, while the CRMSE and error relation clarifies how offsets and error variance contribute to total accuracy.

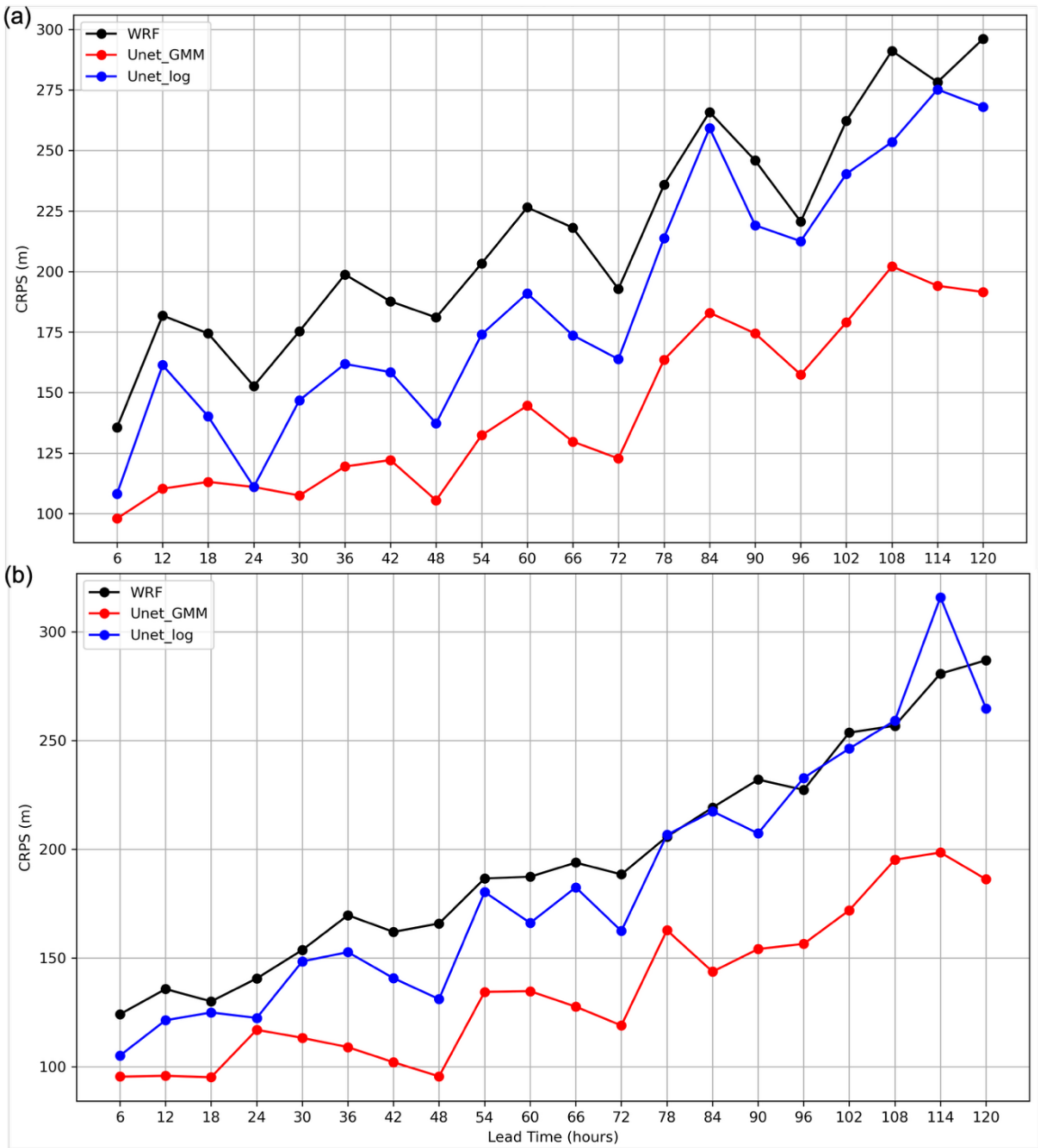

Figure 8. Continuous ranked probability score (CRPS) of WRF FZL, Unet-log and Unet-GMM forecasts with respect to CNRFC observations during (a) December 2016 - March 2017 abs (b) December 2015 - March 2016 over the Yuba-Feather Watershed as a function of lead time (hours).

### *b. Case study: Atmospheric River event over the Western U.S., January 7 –9, 2017*

During January 7–9, 2017, an AR5 event (based on the Ralph et al. 2019 AR Scale) impacted the Western U.S., producing heavy precipitation and widespread hydrological impacts, including flooding and enhanced orographic precipitation in the Sierra Nevada. This event was selected as a case study because ARs are primary drivers of winter precipitation and cause rapid fluctuations in the FZL, which are critical for flood risk management. The event featured an exceptionally strong moisture plume originating from the subtropical Pacific Ocean, which contributed to widespread hydrological impacts, including flooding and enhanced orographic precipitation in the Sierra Nevada. Integrated vapor transport (IVT) values exceeding 1000 kg $m^{-1}$ $s^{-1}$, along with integrated water vapor (IWV) exceeding 30 mm made landfall over central California (Figure 9a-b). Accumulated precipitation during this event exceeded 200 mm over the Sierra Nevada and over 100 mm over the Coastal Ranges of Northern California (Figure 9c). The complexity and operational significance of this AR event make it an ideal case study for evaluating the performance of Unet postprocessing models in improving FZL forecasts under extreme hydrometeorological conditions.

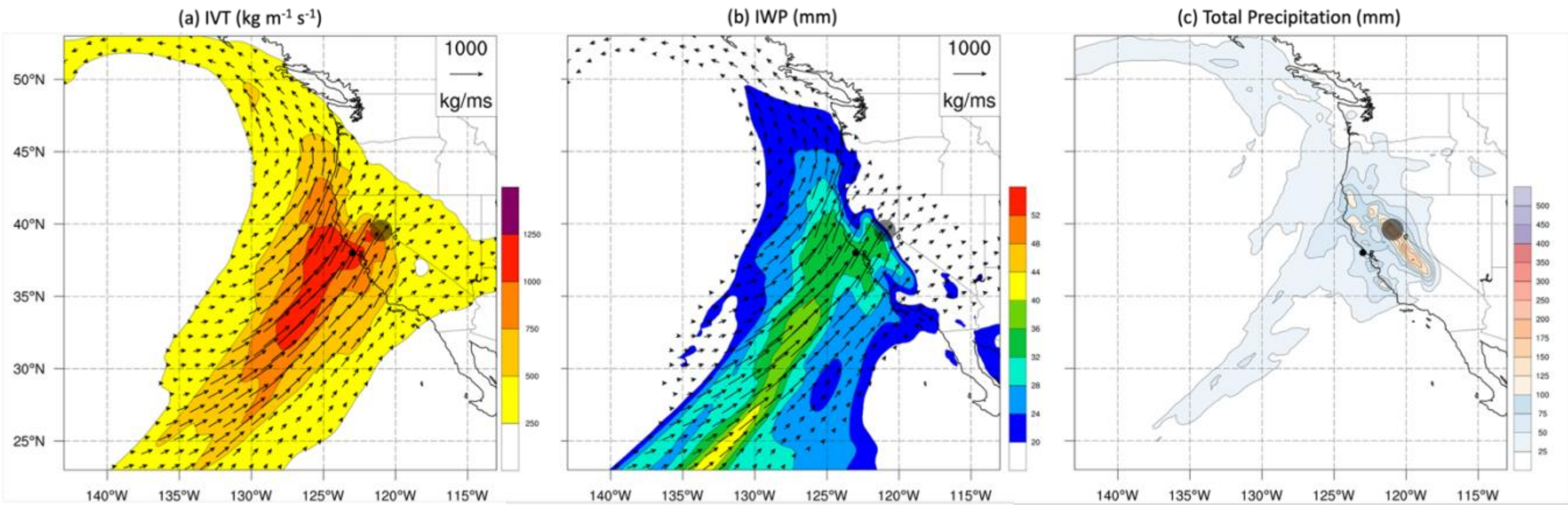

Figure 9. ECMWF ERA5 IVT (kg $m^{-1}$ $s^{-1}$) with IVT vectors (a), IWV (mm) with IVT vectors (b) valid 15 UTC January 8, 2017, and total precipitation (mm) (c) for the AR5 event from 06 UTC January 7 – 08 UTC January 9, 2017. The case study area is indicated by a gray circle.

To assess the performance of the postprocessing models during the AR5 event the spatial distribution of RMSE and correlation between the WRF, Unet-GMM, and Unet-log FZL forecasts and CNRFC observations for lead times of 24–48 hours (Day 2) and 48–72 hours (Day 3) were analyzed (Figure 10). The WRF model exhibits RMSE values exceeding 200 m for Day 2, particularly in the northwestern regions of the basin (Figure 10a). The Unet-GMM (Figure 10b) reduces errors across most of the study region with RMSE values dropping below 150 m, outperforming Unet-log (Figure 10c) for this lead time. For lead time Day 3 (Figure 10d–f), RMSE overall increases across all models, as expected, due to forecast degradation at longer lead times. The WRF model exhibits RMSE values exceeding 550 m in most of the regions, particularly in the lower elevations (less than ~1500 m). The Unet-log and Unet-GMM models similarly lower RMSE values to below 430 m over this region. The Unet-GMM model exhibits slightly higher RMSE values than the Unet-log model but still improves upon the baseline WRF forecast. For Day 2, the spatial structure of correlation differs among the models. WRF shows a modest correlation (Figure 11a), less than 0.73 in many parts of the basin, particularly in areas where RMSE is highest. In contrast, Unet-GMM (Figure 11b) achieves relatively robust correlation values, exceeding 0.78 across most of the watershed, while Unet-log (Figure 11c) demonstrates improved agreement with observations over the southeast and northwest parts of the watershed when compared to WRF. The correlation maps do not show any specific spatial patterns, while forecast skill deteriorates with increasing lead time. For Day 3 (Figure 11d–f), Unet-log maintains relatively high correlations (~0.75–0.90) across much of the watershed, though performance in the southeast and northwest is slightly reduced compared to Day 2. Unet-GMM also exhibits a different pattern from Day 2, with more uniform but slightly lower correlations, remaining around 0.75 across most areas.

Overall, the Unet postprocessing models reduce RMSE across all lead times, with error reductions exceeding 120 m in most areas compared to WRF while maintaining/improving correlation values, particularly at longer lead times. The spatial distribution of improvements varies during this AR5 event, with greater RMSE reductions in western high-elevation regions during Day 2, whereas low-elevation western areas show stronger improvements at longer lead times (Day 3) in comparison to WRF forecasts. As shown in Figure 9, the northern Sierra Nevada, Yuba-Feather watershed included, experiencing the highest IVT and precipitation during the AR event, coincides with areas where the baseline WRF model exhibits the largest forecast errors and low correlations with ground truth (see Figure 10a, d and 11a, d). Appendix C shows sample FZL forecasts for this event from CNRFC, WRF, and the postprocessing Unet-GMM model. These results reinforce the capability of Unet postprocessing to enhance deterministic weather model forecasts, particularly in the context of extreme atmospheric river events where accurate freezing level predictions are critical for hydrological forecasting and flood risk management. Additionally, a sample independent verification against ground-based micro rain radar profilers has been shown in Appendix D to further support these improvements, showing higher detection skills and reduced false alarms compared to raw WRF forecasts. The next section evaluates model performance across different elevation bands to assess the consistency of improvements across varying topography.

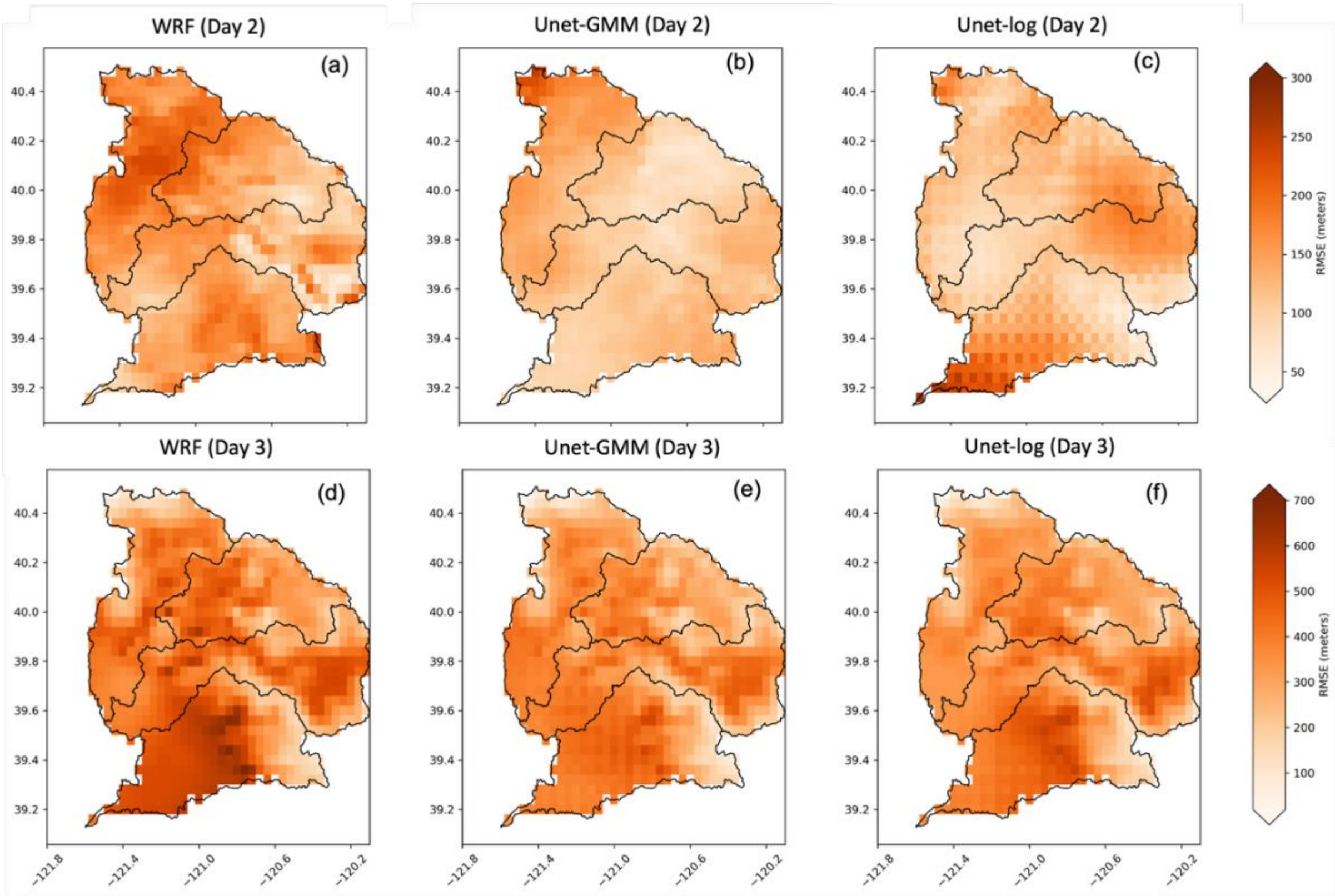

Figure 10. FZL forecasts RMSE for Day 2 lead time (24 to 48 h) (a-c) and Day 3 lead time (48 to 72 h) (d-f) for the case study (initialized on January 7, 2017) for (a and d) WRF, (b and e) Unet-GMM, and (c and f) Unet-log against observations from CNRFC.

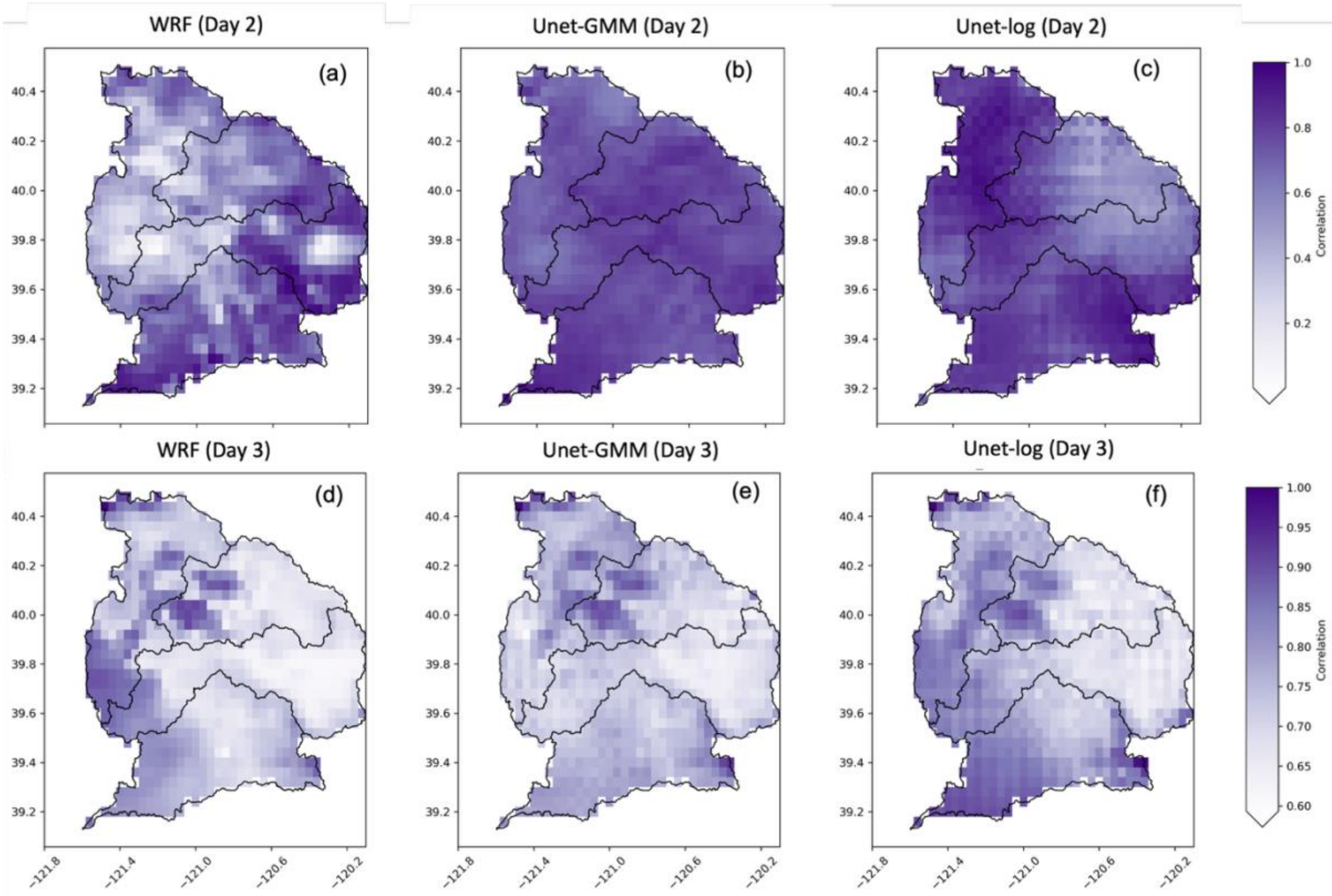

Figure 11. Same as Figure 10, but for the correlation coefficient.

### *c. Performance of Unet postprocessed models across different elevational bands*

To evaluate the effectiveness of the Unet-based postprocessing models in varying topographic conditions, we assessed their performance across five elevation bands within the Yuba-Feather watershed. 3 These analyses were conducted for the winter seasons of 2016-17 and 2015-16. Grid points corresponding to each elevation band were selected, and the results were aggregated, with the number of grid points annotated in the corresponding plots. The panels are organized from the lowest elevation band at the bottom to the highest at the top to facilitate a direct comparison of model performance across terrain levels.

Across all elevation bands, a general decline in correlation is observed with increasing lead time, consistent with the expected loss of forecast skill. The Unet-GMM and Unet-log models demonstrate improved correlation values relative to WRF across all lead times (6–120 hours), with the performance gap between WRF and the Unet-based models remaining consistent. At shorter lead times (6–48 hours), the Unet models achieve high correlations exceeding 0.80, comparable to and better than WRF. However, the performance gap becomes more pronounced beyond 48 hours. By 120 hours, the correlation for WRF drops below 0.70 in high-elevation bands, while Unet-based models maintain values above 0.75. Notably, in high-elevation regions (>1500 m), Unet models exhibit the most substantial improvements in skill, with correlation values ranging from 0.77 to 0.95, depending on lead time. In contrast, at lower elevations (<1500 m), the difference in RMSE and correlation between Unet and WRF is narrower, with Unet models achieving correlation values between 0.85 and 0.98. Across all elevations, the Unet models consistently reduce RMSE relative to WRF, with the most significant improvements observed in high-elevation bands and longer forecast horizons. At elevations above 1750 m, WRF's RMSE often exceeds 350 m beyond 72 hours, while Unet-GMM and Unet-log reduce RMSE by approximately 50–150 m, with Unet-GMM achieving the lowest values throughout. This demonstrates the added value of the Unet postprocessing framework in capturing terrain-dependent forecast errors and improving skill across a range of conditions.

The results for the 2015–2016 winter season exhibit similar trends, albeit with a smaller performance gap between the raw WRF forecasts and the postprocessed outputs from the Unet models (Figure 12). While both Unet models still enhance forecast skill, their advantage over WRF is slightly less pronounced than in the following winter (2016–2017), particularly in the mid and high-elevation bands (>1500 m). Interestingly, for the 2015–2016 season, the correlation values remain above 0.8, while the RMSE values do not fluctuate significantly across different elevation bands. This finding suggests that the performance of the two postprocessing models is influenced by year-to-year variability, synoptic conditions, and seasonal characteristics. However, across both seasons, the Unet postprocessing models consistently improve correlation and reduce RMSE compared to WRF, demonstrating their robustness in enhancing FZL forecasts across diverse topographic conditions and across different winter seasons.

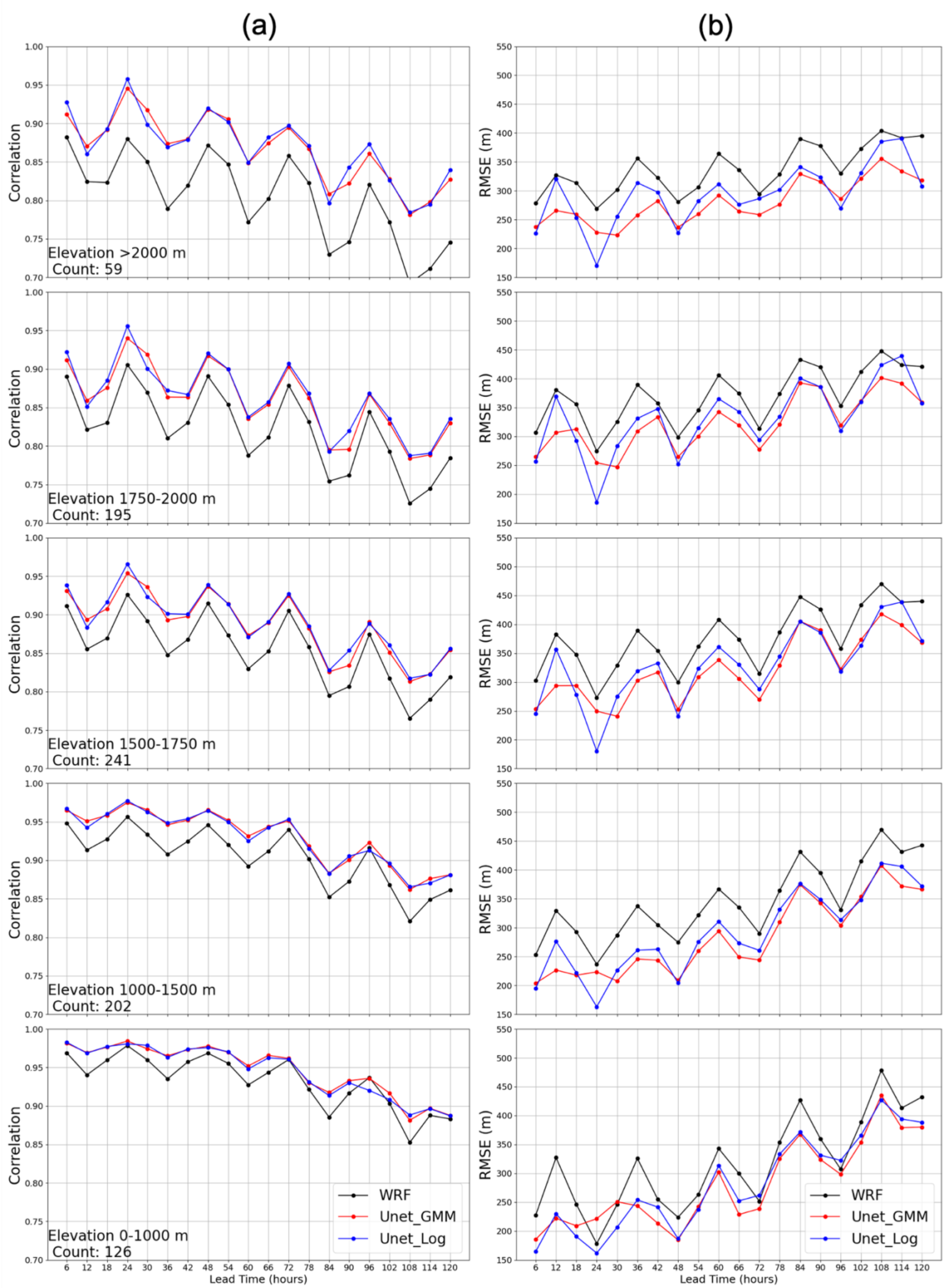

(a)
(b)
Correlation
RMSE (m)
Elevation >2000 m
Count: 59
Elevation 1750-2000 m
Count: 195
Elevation 1500-1750 m
Count: 241
Elevation 1000-1500 m
Count: 202
Elevation 0-1000 m
Count: 126
WRF
Unet_GMM
Unet_Log
Lead Time (hours)

Figure 12. Performance of 6-hourly WRF FZL and two Unet postprocessing forecasts over different elevation ranges during December 2016 - March 2017: (a) correlation coefficient and (b) RMSE versus CNRFC observations as a function of lead time (hours).

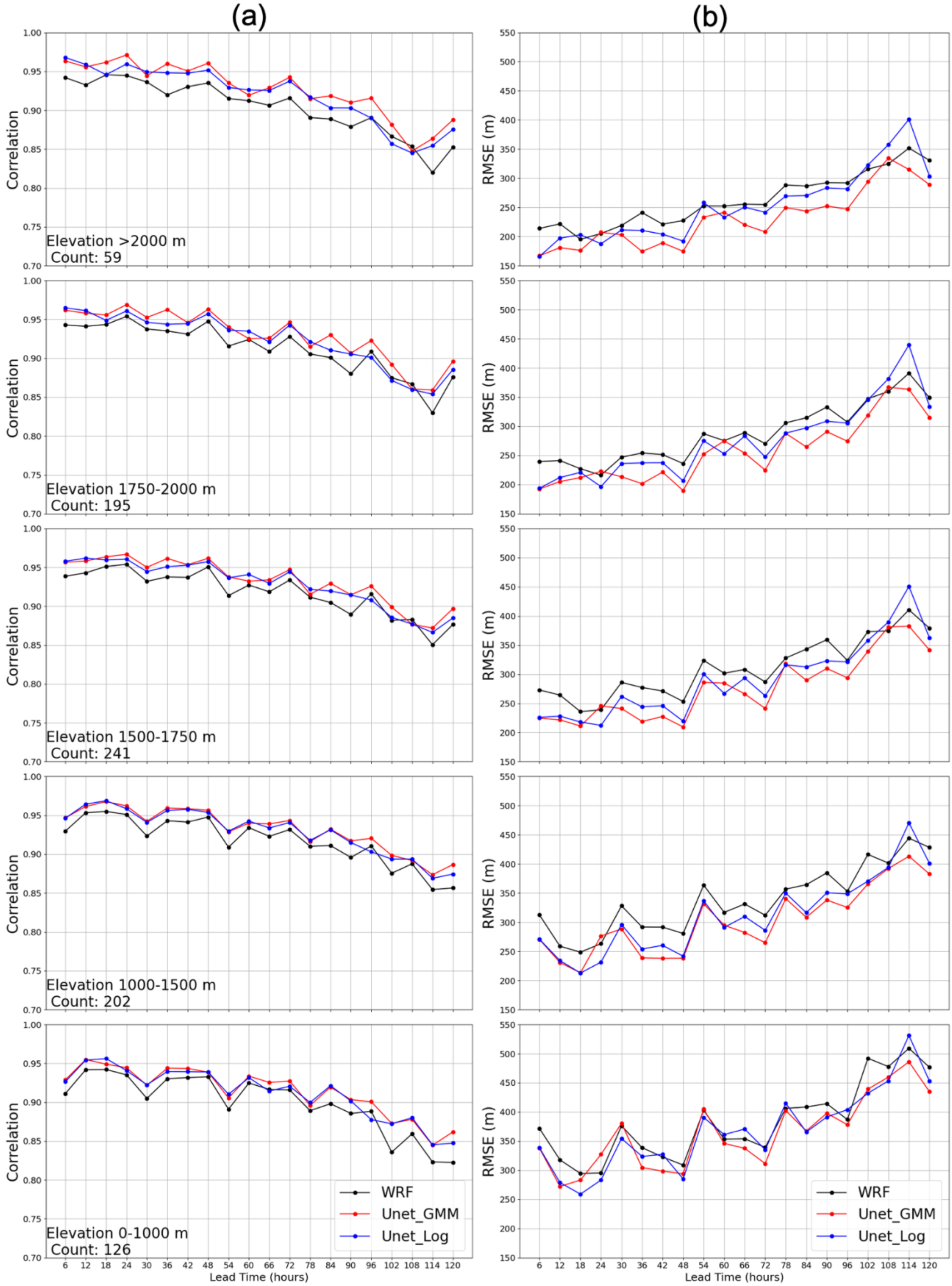
(a)
(b)
Correlation
RMSE (m)
Elevation >2000 m
Count: 59
Elevation 1750-2000 m
Count: 195
Elevation 1500-1750 m
Count: 241
Elevation 1000-1500 m
Count: 202
Elevation 0-1000 m
Count: 126
WRF
Unet_GMM
Unet_Log
Lead Time (hours)

Figure 13. Same as Figure 12 but for winter 2015-2016.

## 5. Conclusions and Future Directions

6This enhancement persists across different topographic regions, in an AR case study, and at forecast lead times up to 5 days, underscoring the value of advanced machine learning frameworks for weather prediction in complex terrain. This postprocessing method can be extended to other watersheds in the Sierra Nevada and beyond. The methodology utilizes the Unet architecture to leverage spatial features in meteorological fields while using Log-Cosh Error and Gaussian Mixture Model (GMM) loss functions, allowing the postprocessing framework to capture a broader range of error distributions than a conventional RMSE-based approach.

Results from the 2016–2017 wet winter show substantial reductions in CRMSE (10–20%) and CRPS (10–35%) and increases in correlation relative to West-WRF, particularly in higher-elevation areas and beyond 72-hour lead times, when raw forecast error increases. Although the drier winter of 2015–2016 exhibits less dramatic overall improvements, Unet-GMM and Unet-Log still consistently improve upon the raw West-WRF, highlighting the robustness of deep learning postprocessing in different precipitation regimes. A case study of a high-impact atmospheric river (AR5; Ralph et al. 2019) event from January 7–9, 2017, demonstrates that Unet-based models effectively capture complex FZL fluctuations and reduce forecast errors, even under extreme moisture transport and intense precipitation. Evaluations across five elevational bands reveal that deep learning corrections are especially beneficial at altitudes above 1500 m, where weather variability, precipitation phase transitions, and terrain-induced processes pose additional challenges for deterministic numerical weather models. Error reductions are also evident at lower elevations and extended lead times, aligning with the broader finding that incorporating ML methods can systematically address biases in physically based models. The Unet models learn spatially varying, flow-conditioned correction to the WRF FZL. Evidence for terrain-dependent learning is the largest CRMSE/CRPS reduction at elevations above 1500 m, where raw WRF biases are largest. Evidence for flow dependence is the growth of skill gains after ~48–72 h and during two test winters as well as the AR case study that synoptic evolution drives rapid FZL shifts. Together, these patterns indicate that the network is removing biases but is also modulating FZL with the evolving large-scale state. Compared to uncorrected West-WRF forecasts, the Unet-based models reduce RMSE, deliver higher correlation coefficients, and more accurately capture the elevations corresponding to the rain–snow transition zone (approximately 1000–2500 m), in alignment with recent studies (Bair et al. 2021; Yang et al. 2022).

From an operational standpoint, by enhancing FZL forecasts, we directly improve the foundation upon which all downstream precipitation-phase and runoff predictions are built. However, it is important to recognize that the 0 °C isotherm does not directly map the surface rain–snow line. In operational forecasts, forecasters address this gap through a "Rain–Snow Elevation" (RSEL) adjustment (an empirically tuned offset typically 300–450 m below the modeled FZL) that represents the mean vertical separation between the atmosphere 0 °C level and the elevation at which hydrometeors transition from snow to rain. Forecasters modulate this RSEL every 6 hours (modeling timestep), directly informing the U.S. National Weather Service SNOW-17 snowpack/rainpack partitioning (Anderson 1973) within the semi-lumped hydrologic models (ie, the Sacramento Soil Moisture Accounting model (SACSMA). Errors in RSEL, thus in the underlying FZL, can translate into significant streamflow forecast errors, potentially consuming tens of percent of flood-pool capacity for a single storm (Sumargo et al. 2020).

While the presented results highlight the promise of deep learning–based postprocessing, they also reflect certain limitations. Our analysis is limited to two evaluation winters, and the framework relies on West-WRF and CNRFC products that may embed systematic uncertainties. Furthermore, the FZL itself can be a proxy for the rain–snow line, and results should be interpreted in an operational context. Encouraged by these findings, future research will investigate how these FZL forecast enhancements are propagated to inflow forecasts and, ultimately, inform reservoir decision-making. Following approaches similar to Sumargo et al. (2020), a quantitative assessment of how FZL error reductions affect reservoir operations will be undertaken—an essential step toward integrating deep learning postprocessing into hydrologic and operational models. Looking ahead, several avenues exist for enhancing the efficacy and robustness of Unet-based postprocessing. These include further development of probabilistic frameworks beyond just two-component GMM, evaluation of ensemble approaches for uncertainty quantification (such as the study by Hu et al. 2023 for precipitation postprocessing), exploring training Unet frameworks for individual watersheds or transfer learning to adapt models to other mountain basins, and incorporating additional atmospheric variables (e.g., humidity profiles, wind speeds) into the input feature space. Such extensions are poised to further boost predictive skill and resilience under a variety of weather conditions.

In addition, we recognize a growing body of literature highlighting the advantages of wet bulb temperature and dew point temperature as diagnostic variables for predicting surface precipitation phase, particularly during mixed-phase events common in complex terrain. Studies by Jennings et al. (2025, 2023, 2018), Froidurot et al. (2014), and Marks et al. (2013) demonstrate that these variables, by capturing critical humidity effects, frequently outperform predictions based solely on the air temperature of 0°C. However, our current study intentionally addresses a different, operationally driven challenge, improving FZL forecasts due to their direct and established use within operational hydrologic forecasting frameworks, such as by CNRFC. Future research will therefore include comparative assessments of our ML-informed FZL forecasts against these humidity-based variables and the operationally implemented RSEL approach. Such comparative analyses will inform potential future adjustments to operational forecasting methodologies, enhancing their capacity to manage uncertainties under diverse hydrometeorological conditions.

Overall, this study offers a promising advancement for applying modern deep learning architectures to resolve long-standing challenges in complex-terrain FZL forecasting. By systematically reducing model errors and their variations, lowering CRPS to capture distributional accuracy, and explicitly acknowledging limitations and future extensions, enhancing spatial coherence, and improving lead-time performance, Unet-based postprocessing demonstrates the potential to pave the way for more accurate and adaptive hydrometeorological services, particularly important in a changing climate where extreme precipitation events are growing in frequency and intensity (Espinoza et al. 2018).

*Acknowledgments.*

This work is supported by the U.S. Army Corps of Engineers (USACE) Forecast Informed Reservoir Operations Phase 3 (USACE W912HZ242) and the Yuba County Water Agency (Award # 37707-2022-30145954). We thank Robert Hatman, Brett Whitin, Tyler Brandt, Kayden Haleakala, Taylor Dixon and Anna Wilson for fruitful discussions and their support in the development of the scheme and manuscript preparations.

*Data Availability Statement.*

The California Nevada River Forecast Center (CNRFC) gridded freezing level observations are publicly available at https://www.cnrfc.noaa.gov/arc_search.php?product=netcdfoz. The details of CNRFC freezing level products are available at https://www.cnrfc.noaa.gov/fzlvl_guidance.php#:~:text=The%20text%20and%20graphical%20products,Freezing%20Levels%20%2D%20Interactive%20Map%20Interface. The Center for Western Weather and Water Extremes (CW3E) West-WRF output is too large to be publicly archived. Access to reforecasts, along with documentation and methods used to support this study, are available from Daniel Steinhoff at CW3E/UCSD (dsteinhoff@ucsd.edu).

## APPENDIX A

### Comparison of Winter Precipitation and Freezing Levels

This appendix presents accumulated precipitation from the European Centre for Medium-Range Weather Forecasts' fifth-generation reanalysis product (ERA5) to illustrate the contrast between the wet and dry winters used in the model test period: December 2015–March 2016 and December 2016–March 2017. These figures support the discussion in the last paragraph of Section 3.3. Figure A1a shows the total accumulated precipitation for the winter of 2015–2016, which was characterized by below-average precipitation despite a strong El Niño event. Figure A1b displays the accumulated precipitation for the winter of 2016–2017, a season noted for record-breaking rainfall across California. To better highlight spatial differences between the two winters, Figure A1c shows the difference in accumulated precipitation (2016–17 minus 2015–16). These figures provide additional context for interpreting model performance under contrasting hydroclimatic conditions.

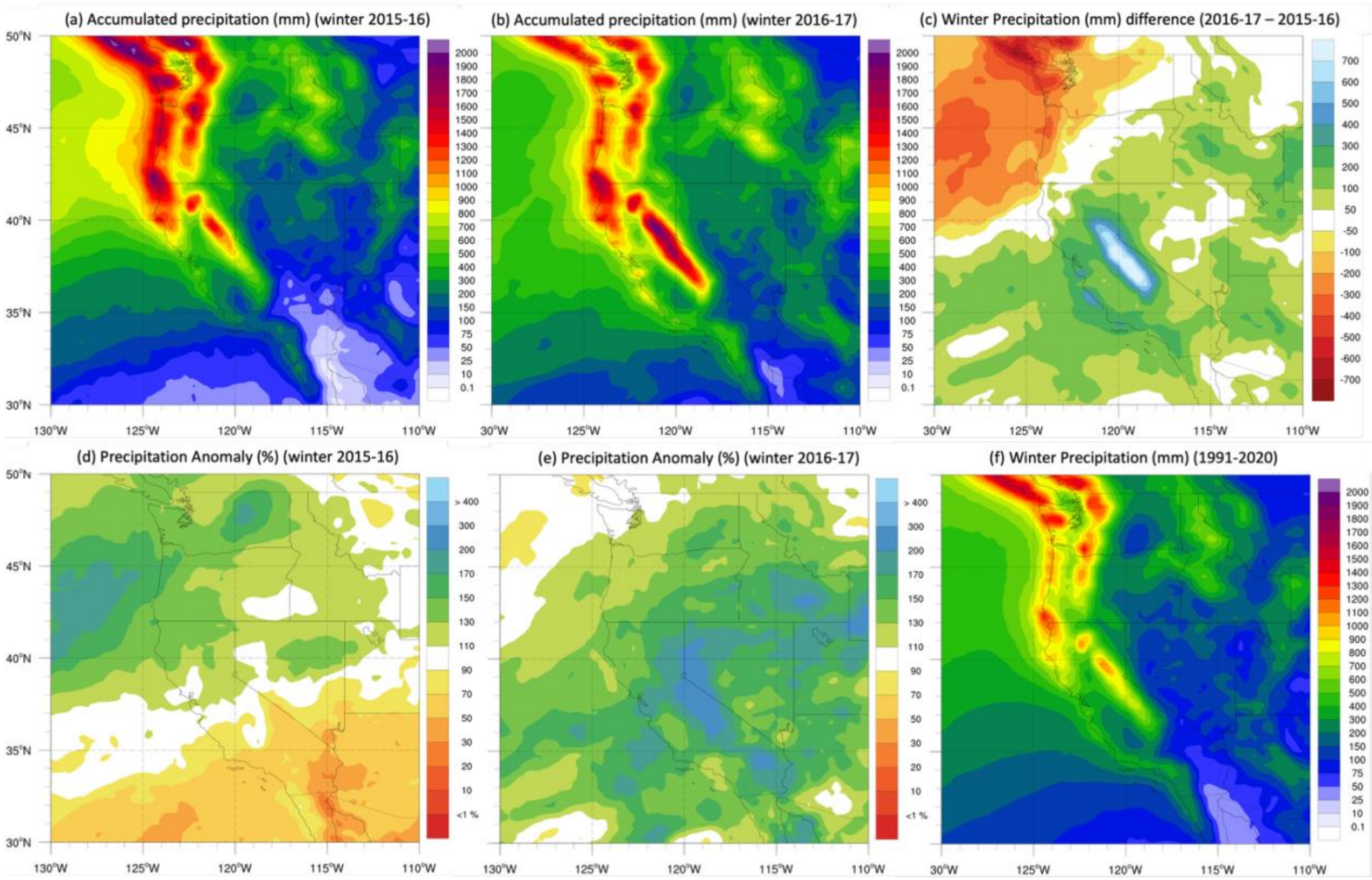

Figure A1. ERA5 accumulated precipitation (mm): (a) from December 1, 2015, to March 31, 2016; (b) from December 1, 2016, to March 31, 2017; (c) difference in accumulated precipitation (mm) between the two test winters (2016–2017 minus 2015–2016); (d) precipitation anomaly for winter 2015–16 relative to the 1991–2020 climatology; (e) precipitation anomaly for winter 2016–17 relative to the 1991–2020 climatology; and (f) climatological mean winter precipitation (December–March) for 1991–2020.

In addition to precipitation patterns, Figure A2 illustrates a sample visualization of CNRFC gridded FZL and modeled FZL from WRF for a winter day in January of 2016. Panel A2a shows the CNRFC FZL (m), while Panel A2b displays the corresponding WRF model FZL (m). This example highlights spatial differences in FZL estimates between the operational observations and the numerical weather prediction model, illustrating the type of discrepancies addressed by the postprocessing approach discussed in the main text.

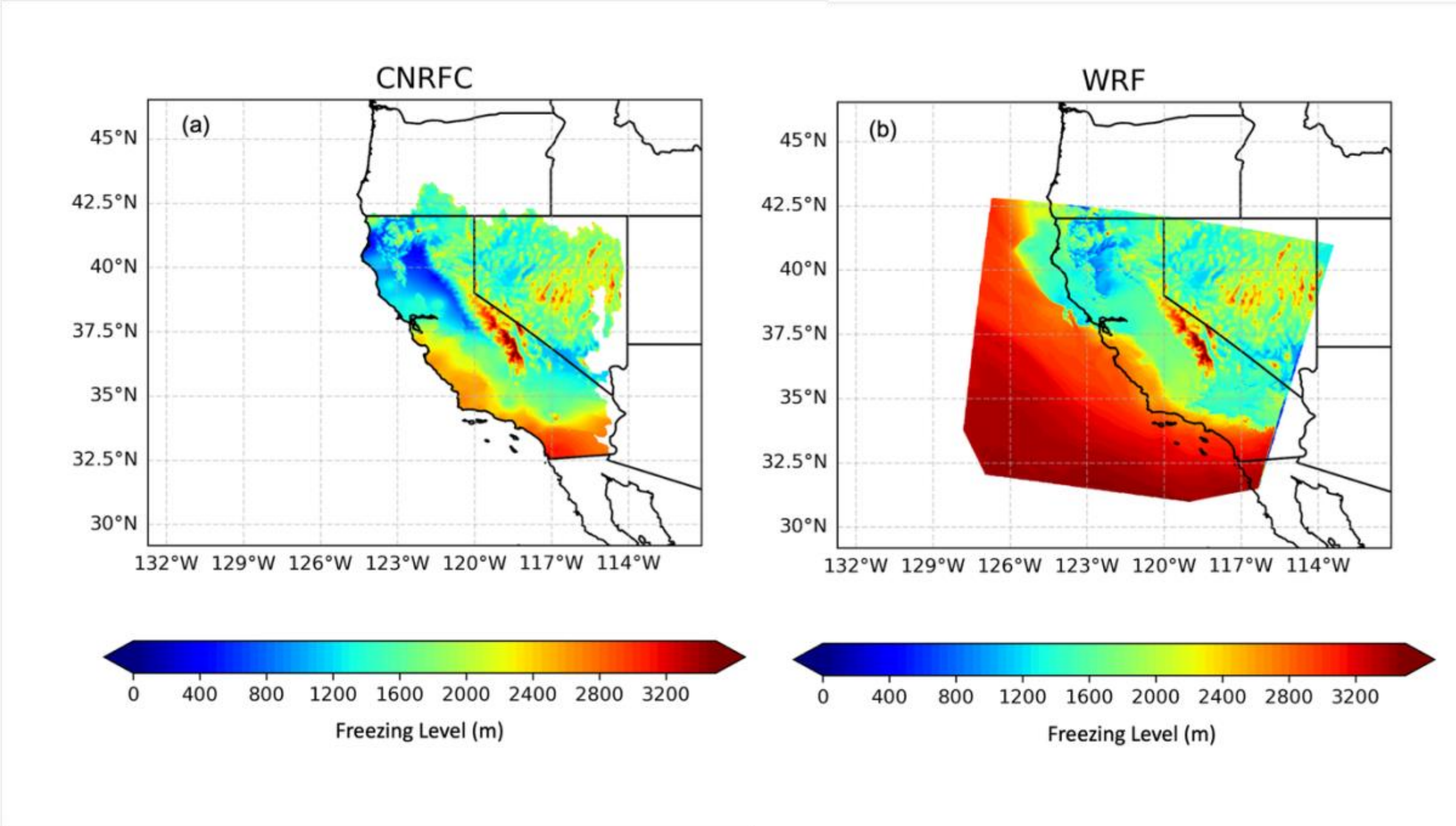

Figure A2. Freezing level (FZL) comparison for daily FZL on January 1, 2016: (a) CNRFC FZL (m); (b) WRF model FZL (m).

# APPENDIX B

**Verification Metrics**

The performance of deterministic postprocessing models is evaluated with Pearson correlation coefficient (higher is better, 1 is perfect), Mean Absolute Error (MAE), root-mean-square error (RMSE) and centered root-mean-square error (CRMSE, lower is better, ideal is 0):

$$\text{Correlation} = \frac{1}{n}\frac{\sum_{i=1}^{n}(P_i-\overline{P})(O_i-\overline{O})}{\sqrt{\sum_{i=1}^{n}(P_i-\overline{P})^2}\sqrt{\sum_{i=1}^{n}(O_i-\overline{O})^2}}$$

$$\text{MAE} = \frac{1}{n}\sum_{i=1}^{n}|P_i - O_i|$$

$$\text{RMSE} = \sqrt{\frac{1}{n}\sum_{i=1}^{n}(P_i - O_i)^2}$$

$$\text{CRMSE} = \sqrt{\frac{1}{n}\sum_{i=1}^{n}\left((P_i-\overline{P})-(O_i-\overline{O})\right)^2}$$

Where $P_i$ represents the $i^{th}$ FZL forecasts, $O_i$ refers to the corresponding FZL observation, with $n$ denoting the total number of samples.

To evaluate the probabilistic postprocessing outcome from Unet-GMM, CRPS (Continuous Ranked Probability Score) is used, that measures the difference between the predicted cumulative distribution function (CDF) and the observed value. For a forecast CDF $F$ and an observation $O$:

$$\mathrm{CRPS}\,(\mathrm{F},\mathrm{O}) = \int_{-\infty}^{\infty} [F(P) - 1\,(\mathrm{P} \geq \mathrm{O})]^2\, dP$$

Where $1\,(\mathrm{P} \geq \mathrm{O})$ is the indicator function (1 if $\mathrm{P} \geq \mathrm{O}$, 0 otherwise). For deterministic forecasts, CRPS generates MAE.

APPENDIX C

**Freezing Level Forecasts for the AR5 Case Study**

This appendix presents a visual comparison of the raw WRF and postprocessed FZL forecasts using Unet-GMM for the AR5 event described in Section 4b. The forecasts shown here were initialized at 0000 UTC on 8 January 2017, during the onset of the strong AR5 event that impacted the western United States from 7–9 January 2017. The panels in Figure C1 illustrate Day 1 (24-h lead) and Day 2 (48-h lead) FZL forecasts from the raw WRF model and the postprocessed model over Yuba-Feather Watershed. This example complements the quantitative evaluation in Section 4b by highlighting spatial differences between the WRF forecasts and their postprocessed values. The postprocessing reduces regional biases and better aligns the forecasted FZL with observed patterns from CNRFC, especially over complex terrain where accurate representation of the FZL is critical for hydrologic modeling and flood risk management.

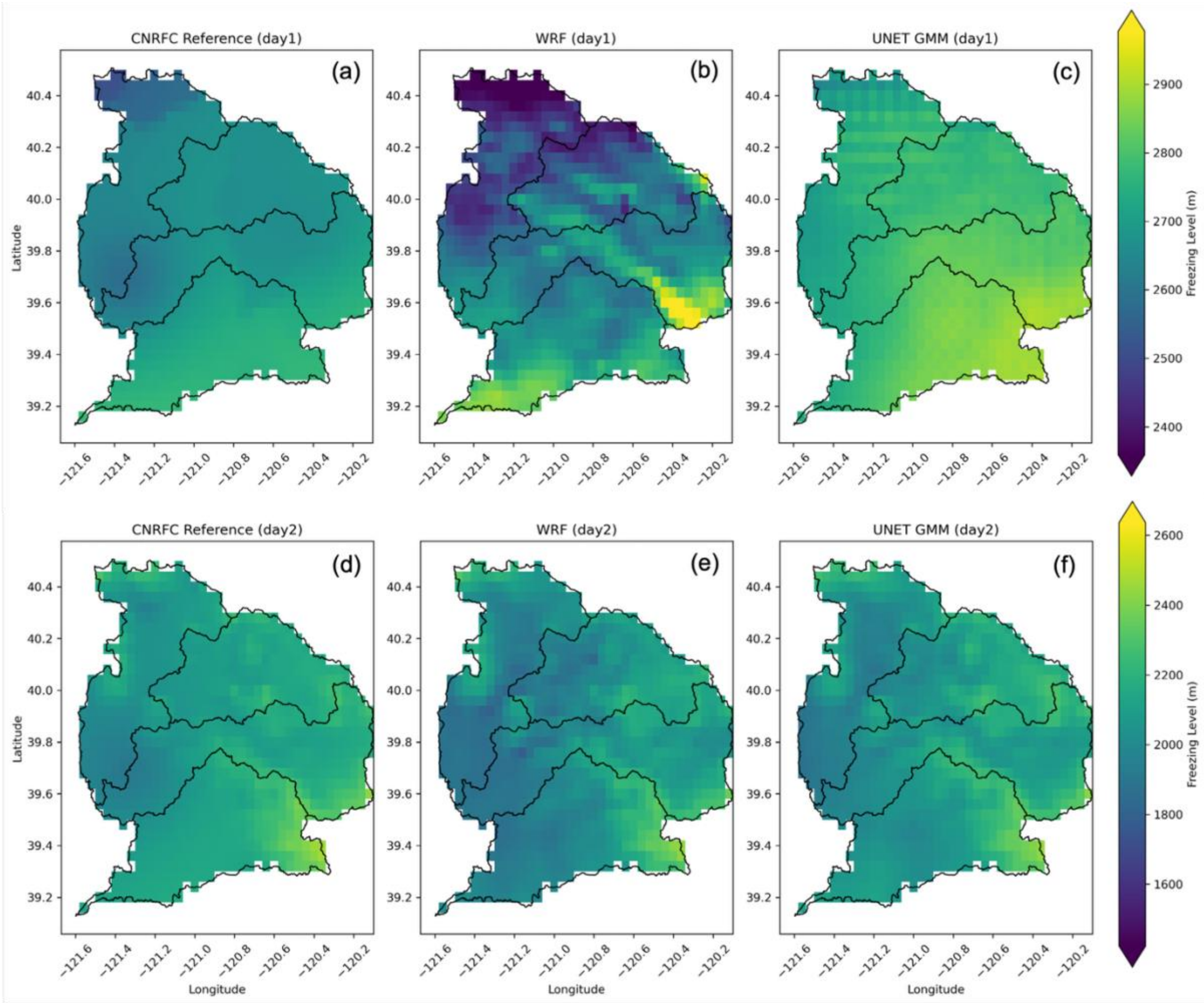

Figure C1. Freezing level values (FZL; m above mean sea level) for the AR5 event initialized at 0000 UTC 8 January 2017: (a) reference FZL from CNRFC for Day 1 (24-h lead), (b) WRF Day 1 forecast, (c) postprocessed Day 1 forecast from the Unet-GMM model, (d) reference FZL from CNRFC for Day 2 (48-h lead), (e) WRF Day 2 forecast, and (f) postprocessed Day 2 forecast from the Unet-GMM model.

## APPENDIX D

### FZL at Bullards Bar Dam (NBB) and Downieville (DLA) profilers

For verification of freezing level (FZL) detection using pointwise ground-based observation, we use two micro rain radar (MRR) observations from the New Bullards Bar Dam (NBB) and Downieville (DLA) profilers in the northern Sierra Nevada. These two sites are within the study area and have minimal missing data. The NBB MRR is located at 39.3964° N, 121.1438° W, at an elevation of 632 m, while the DLA MRR is situated at 39.5633° N, 120.8243° W, at an elevation of 906 m. Both profilers operate at 15-minute temporal resolution with 100 m vertical resolution. The MRRs are K-band (24 GHz) Doppler weather radars that employ frequency-modulated continuous-wave (FMCW) signals to measure radar reflectivity and Doppler vertical velocity, from which the altitude of the bright band, a proxy for the melting level, is derived using the White et al. (2002) algorithm. We evaluated the improvement of FZL detection from WRF forecasts after postprocessing with our Unet-GMM approach during the winter season (December–March) of 2022–2023. Results for this independent verification from NBB profiler are shown in Figure D1, and results for the DLA profiler are presented in Figure D2.

The hit is defined as the forecasted FZL is within ±200 m of the observed FZL from the profiler, while Miss is when an observed FZL occurs, but the forecasted FZL is more than 200 m away from the observed value. False Alarm is also defined as the forecasted FZL indicates detection within ±200 m of the site's grid elevation, but the observation shows no such detection. The Critical Success Index (CSI) is defined as:

$$\mathrm{CSI} = \frac{\mathrm{hits}}{\mathrm{hits} + \mathrm{misses} + \mathrm{false\ alarms}}$$

The results indicate that Unet-GMM improves FZL detection relative to the raw WRF forecasts. The number of hits and misses remains similar between WRF and Unet-GMM, but the postprocessed forecasts show a clear reduction in false alarms, leading to higher CSI values for lead times from Day 2 through Day 4. This improvement is consistent across both profiler sites and demonstrates the ability of the postprocessing approach to reduce spurious detections while preserving correct ones.

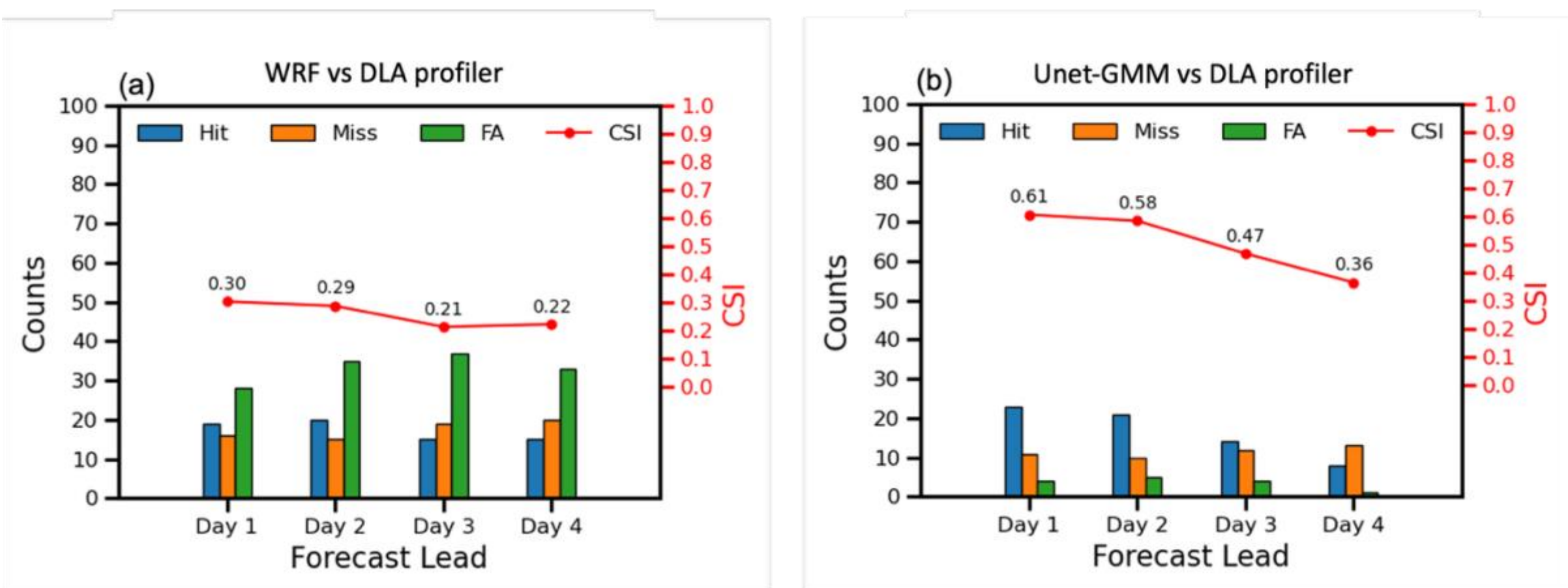

Figure D1. Comparison of FZL detection at Downieville (DLA) MRR profiler during winter (December–March) 2022–2023: (a) raw WRF FZL versus profiler, and (b) postprocessed UNet-GMM FZL versus profiler. Bars indicate the number of hits (blue), false alarms (green), and misses (orange), with counts shown on the left y-axis. The red line shows the CSI values (right y-axis) for forecast lead times from Day 1 to Day 4.

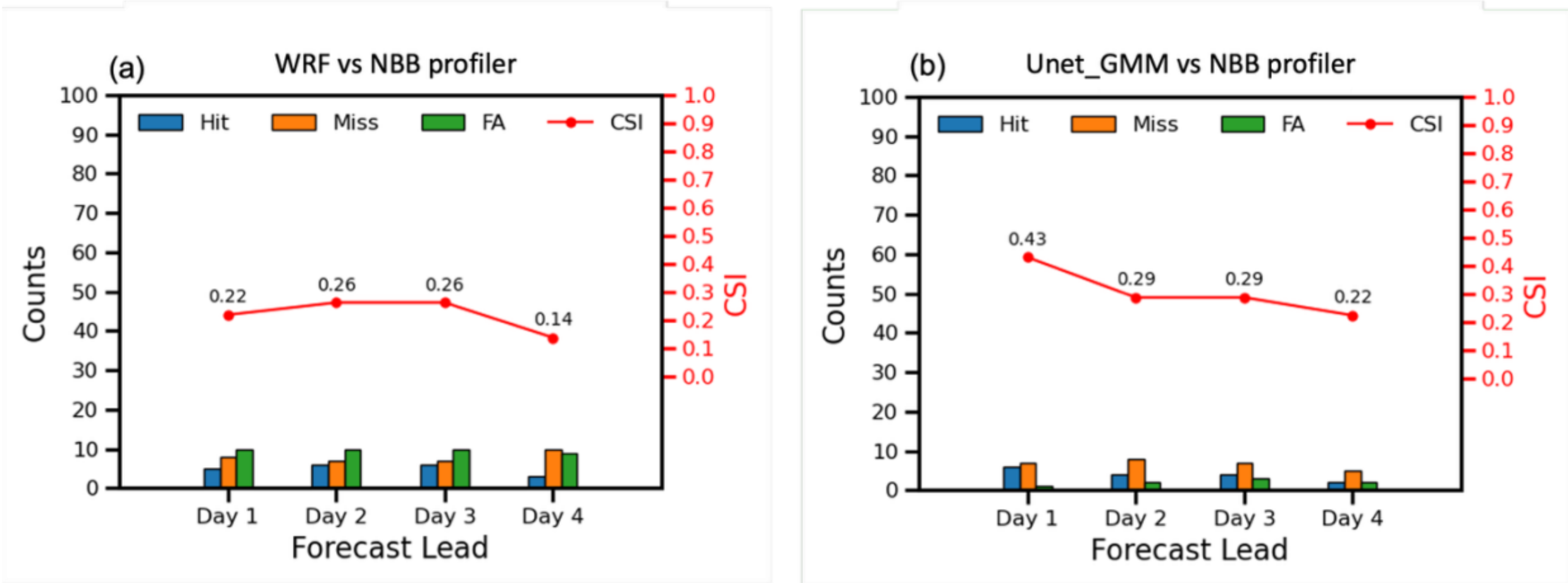

Figure D2. Same as Figure D1, but for New Bullards Bar Dam (NBB) MRR profiler.